\documentclass[10pt,conference]{IEEEtran}

\usepackage[numbers,sort&compress]{natbib}

\usepackage{amsmath,amssymb,amsfonts}
\usepackage{pifont}

\usepackage{algorithm}
\usepackage{algorithmic}
\usepackage{booktabs}
\usepackage{graphicx}
\usepackage{textcomp}
\usepackage[table,dvipsnames]{xcolor}
\usepackage{fancyhdr}
\usepackage{hyperref}
\usepackage{tabularx}

\usepackage{subcaption}  % modern way to handle subfigures
\usepackage{caption}
\usepackage{wrapfig}
\usepackage{enumitem}
\usepackage{xspace}
\usepackage{needspace}
\usepackage{minted}
\usepackage{microtype}
\usepackage{multirow}
\usepackage{threeparttable}

\usepackage[export]{adjustbox}
\usepackage{cuted}
\usepackage{soul}

\usepackage{tikz}
\newcommand*\circled[1]{\tikz[baseline=(char.base)]{
            \node[shape=circle,draw,inner sep=1pt] (char) {\small#1};}}

\newcommand{\figref}[1]{Fig.~\ref{#1}}

\newcommand{\design}{SPARTA\xspace}

\newcommand{\zyz}[1]{}
\newcommand{\rev}[1]{#1}
\colorlet{keyhl}{yellow!55}\sethlcolor{keyhl}
\soulregister{\cite}{7}\soulregister{\ref}{7}\soulregister{\figref}{7}\soulregister{\design}{0}
\newcommand{\key}[1]{#1}
\newcommand{\hzy}[1]{}

\title{\design: A Sparse Ternary LLM Accelerator with Adaptive Attention and Unified Datapath for Long-Context Edge Inference}
\usepackage[pass]{geometry}
\newif\ifwithcover\withcoverfalse
\ifwithcover
\AtBeginDocument{\newgeometry{margin=0.6in}\thispagestyle{empty}% Cover letter body, shared by cover_page.tex (standalone) and main.tex (\withcovertrue).
\definecolor{rA}{HTML}{FF1744}
\definecolor{rB}{HTML}{00A6ED}
\definecolor{rC}{HTML}{00C853}
\definecolor{rD}{HTML}{FF8F00}
\definecolor{rE}{HTML}{D500F9}
\newcommand{\RA}[1]{\textcolor{rA}{\textbf{A-#1}}}
\newcommand{\RB}[1]{\textcolor{rB}{\textbf{B-#1}}}
\newcommand{\RC}[1]{\textcolor{rC}{\textbf{C-#1}}}
\newcommand{\RD}[1]{\textcolor{rD}{\textbf{D-#1}}}
\newcommand{\RE}[1]{\textcolor{rE}{\textbf{E-#1}}}
\newcommand{\Sec}[2]{\par\vspace{6pt}\noindent{\normalsize\textbf{#1\enspace #2}}\par\vspace{1pt}}
\newcommand{\Q}[3]{\par\vspace{1pt}\noindent\textbf{#1}\enspace #2\enspace\textbf{#3.}\enspace}
{\small
\noindent{\large\textbf{Paper \#1296 --- Summary of Revisions}}\par\vspace{3pt}
\noindent We thank the reviewers for their constructive comments, which helped us strengthen the evaluation. Sections follow the meta-review points in order. Reviewers \textcolor{rA}{\textbf{A}}--\textcolor{rE}{\textbf{E}} are color-coded; tags give the reviewer and item (Q: question for revision, W: weakness; for D, the commented figure or section), and related items are merged. Changes are marked in \textcolor{blue}{blue}, and the most important ones are also {\setlength{\fboxsep}{0.5pt}\colorbox{yellow!55}{highlighted in yellow}}; figure, table, and section numbers refer to the revised paper.

\Sec{1}{Differentiation from prior work}
\Q{1.1}{\RA{Q1} \RA{Q6}}{Bit-level sparsity and VitaLLM} \design's novelty is a single integer datapath shared by ternary linear layers and bit-serial attention, which the cited bit-level and ternary designs do not provide. Unlike Bit-Pragmatic, Bit-Tactical, and Laconic, each set Q/P bit drives a select-add of a whole K/V vector, and unlike BitMoD, operands stay integer (Sec.~VI). VitaLLM, now in Table~I and Secs.~I and VI, runs ternary layers on TINT cores but attention on a separate BoothFlex core, so it cannot reallocate compute as the workload shifts.

\Q{1.2}{\RB{W1} \RD{bullet\,(3)}}{Scope of two-stage selection} Two-stage selection is not ternary-specific, but ternary LLMs are where it matters most and is hardest to support: their tiny weights leave the KV cache dominating decode traffic (Fig.~1(b)), while ternary accelerators rely on separate engines for multi-bit attention. The novelty is not the selection policy, which follows Quest and Twilight (Sec.~III-C), but running it on the ternary fabric itself: bit-serial QK/PV reuse the select-add datapath, and dedicated engines make selection cheap.

\Sec{2}{Experiments comparing the unified datapath vs heterogeneous or dual-datapath}
\Q{2.1}{\RA{Q5} \RC{Q1} \RD{C2} \RE{W1}}{Dual-datapath baseline} A unified datapath is the right choice for edge inference: it serves both linear and attention layers with higher energy and area efficiency in a far more compact chip, which is exactly what edge devices are constrained by. Against TENET, whose separate linear (STL) and attention (HP) cores represent heterogeneous designs, \design delivers $1.4\times$ the energy and $2.2\times$ the area efficiency in about 1/7 of the silicon (Fig.~22, Table~V). Our PE also beats the linear PEs of TeLLMe, FIGLUT, and TENET in throughput per area (Fig.~4). PADE and SpAtten accelerate attention only, so for the end-to-end comparison we add MCBP to Fig.~22.

\Sec{3, 4}{HBM traffic reduction vs tile-granularity; HBM energy inclusion in efficiency metrics}
\Q{3.1}{\RA{Q2}}{HBM transfer granularity} Our traffic savings hold at real HBM transfer granularity: replayed in Ramulator2, the transferred bytes are only 1.4--1.7\% above the analytical estimate (Sec.~V-C, Fig.~17).

\Q{4.1}{\RE{Q2}}{Energy accounting and simulator} Energy is accounted consistently across designs, and the scope of each measurement is now stated explicitly. In Fig.~14, \design's energy includes HBM, whereas CPU energy (RAPL package) excludes DRAM and Jetson energy covers its module rails. Fig.~22 counts on-chip and HBM dynamic energy for all designs under one HBM2 model, while Table~V lists on-chip power only because it compares published chip-level metrics of designs with different memory systems.

\Sec{5}{Experimental metrics and missing tests}
\Q{5.1}{\RB{W3} \RD{Tab.\,III} \RE{W5}}{Quality loss} \design stays close to dense attention on most tasks: Table~III summarizes per-task scores with a geometric mean, Sec.~V-B explains the larger losses, and Sec.~I now states a task-dependent loss.

\Q{5.2}{\RD{III.C} \RE{W5}}{Retrieval tests} Retrieval, the setting where sparse attention usually fails first, depends on the model: BitCPM-3B largely retains it, while Bonsai-4B exposes the limit of a fixed page budget. On BitCPM-3B, passkey passes at every depth and length, and four LongBench retrieval and code tasks lose 1.5 points on average. Bonsai-4B passes passkey up to 16K (16/20 at 32K) and drops from 73.7 to 52.4 on passage retrieval (Sec.~V-B), which a per-task or per-layer page budget can address in future work.

\Sec{6}{Missing baselines and unclear modeling of existing ones}
\Q{6.1}{\RC{Q2} \RE{Q3}}{Twilight} \design matches Twilight's selection quality while computing the Top-$p$ threshold more efficiently in hardware. On the same fabric, Twilight's binary-search threshold slows attention by 2.0--3.0$\times$ (Sec.~V-D), while our histogram stays within 0.10 PPL of Twilight's exact selection (Table~IV). The CPU/GPU baselines use dense attention because \texttt{llama.cpp} lacks Quest and Twilight.

\Q{6.2}{\RE{Q1}}{Baseline modeling} Our baseline models are faithful and, if anything, favor the baselines. We reproduce their linear-layer cores in RTL at the published widths and synthesize them in our 28-nm flow; TENET's HP cores use its published size with calibrated power, and attention beyond the published lengths runs densely on the shared HBM2 (Sec.~V-A). Our FIGLUT and TeLLMe models run faster than published, and TENET's STL array matches its published throughput.

\Sec{7}{Unclear implementation details}
\Q{7.1}{\RD{Q1} \RD{Q3}}{Merge, histogram, and partial pages} The selection engines have a fixed, modest footprint, and every buffer they need is already counted in our area and performance results. $k$ is capped at 512 pages (Sec.~III-C), the bitonic merge and 64-bin histogram use fixed buffers (Sec.~IV-C), and only the 2048-entry page-score SRAM scales with context. Padding and unfilled tail slots are never selected.

\Q{7.2}{\RA{Q4} \RD{Q2}}{Stage cycles and latency hiding} Selection runs concurrently with the new token's K/V generation, and our results charge every remaining stall: Top-$p$ covers V generation and Top-$k$ covers K generation except for Bonsai at 8K (Fig.~16).

\Q{7.3}{\RB{Q2} \RD{III.C}}{Candidate-set Top-$p$} Applying Top-$p$ within the Top-$k$ candidates is deliberate: it is always satisfiable and bounds hardware work. $p$ is normalized over the candidates' mass, so no pages are added even when they hold less than $p$ of the full mass (Sec.~III-C); Top-$p$ changes perplexity by at most 0.23 (Table~IV), so most loss comes from the page budget.

\Q{7.4}{\RB{W2} \RB{Q1}}{Quantization} The gains do not come from quantization: dense and sparse attention use the same INT4 KV. Sec.~V-A specifies all quantizers, and with an INT4-KV Jetson baseline (Sec.~V-C) our gains grow further.

\Q{7.5}{\RD{Fig.\,7}}{Query precision and Fig.~7} The datapath handles INT8 queries directly (Sec.~V-A), and Fig.~7(c) now labels the operands as in the pseudocode.

\Sec{8}{Missing GQA/MQA and batch-size analysis}
\Q{8.1}{\RB{Q4}}{GQA/MQA} \design supports GQA and MQA natively by sharing one selection per KV group, which real GQA models require. Sec.~III-C adds group-shared selection, with MQA as the single-group case; per-head selection would read $3.1\times$ the dense traffic.

\Q{8.2}{\RB{Q3}}{Batch size} Larger batches make sparsity even more valuable. Decode linear layers become SpMM, and from batch 1 to 8 the sparse-over-dense gain grows from $1.5\times$ to $2.2\times$ at 8K and from $2.3\times$ to $3.4\times$ at 32K.

\Sec{9}{Design generality and robustness}
\Q{9.1}{\RA{Q3} \RE{W5}}{Fig.~3(b) and adaptive budget} Attention needs vary widely across heads and inputs, which is exactly why \design pairs a bounded Top-$k$ with an adaptive Top-$p$. At 32K, the tokens holding 90\% of the mass vary 35--44$\times$ from p5 to p95 (setup in the Fig.~3(b) caption), and Top-$p$ keeps only 2.9--10.9\% of the candidates (Table~IV).

\Q{9.2}{\RE{W5}}{Generality} The benefits carry over across ternary model families and sizes. KV traffic drops by 4.1--4.8$\times$ for BitCPM-3B and Bonsai-4B, and Fig.~14 adds BitCPM-3B/8B.

\Q{9.3}{\RC{Q3}}{Weight sparsity} \design degrades gracefully when ternary models are less sparse. From 0\% to 60\% zero weights, prefill and decode stay within 0.69--1.20$\times$ of the 40\% point, where the bitmap matches 5-trit packing.

\Q{9.4}{\RE{Q4}}{Chunked prefill} Page Top-$k$ extends naturally to chunked prefill by using each chunk's queries, which we leave to future work.

\Sec{10}{Writing issues}
\Q{10.1}{\RD{W} \RD{Fig.\,1}}{Prior work and Fig.~1} Sec.~II now credits each prior design for its goals and tradeoffs and corrects PADE as predictor-free. Fig.~1 defines intelligence retention against the full-precision base models, and Fig.~1(b) shows traffic, not footprint.

\Q{}{}{Other changes (not highlighted)} Numbers in the abstract and Secs.~I, IV-B, and V-B--V-F are updated with the GQA-aware simulator.
\par}
\clearpage\restoregeometry\setcounter{page}{1}}
\fi

\author{
  \IEEEauthorblockN{Yanzhen Zhou\IEEEauthorrefmark{1}, Zhengyu Han\IEEEauthorrefmark{1}, Qunyou Liu\IEEEauthorrefmark{2}, Zheyu Yan\IEEEauthorrefmark{1}, Weihong Xu\IEEEauthorrefmark{1}\IEEEauthorrefmark{3}, Cheng Zhuo\IEEEauthorrefmark{1}}
  \IEEEauthorblockA{\IEEEauthorrefmark{1}Zhejiang University \qquad \IEEEauthorrefmark{2}Veloxis AI \qquad \IEEEauthorrefmark{3}Corresponding author \\
    \{yanzhenzhou, zhengyu.han, zyan2, wh.xu, czhuo\}@zju.edu.cn, qy.liu@veloxis-tech.com}
}

\begin{document}

\maketitle
\thispagestyle{plain}
\pagestyle{plain}

%%%%%%%%%%%%%%%%%%%%%%%%%%%%%%%%%%%%%%%%
%%%%%%%% -- PAPER CONTENT STARTS -- %%%%%%%%%

\begin{abstract}
  Large language models (LLMs) are increasingly deployed on edge devices, where limited memory capacity and bandwidth constrain both model size and inference efficiency. Recent ternary LLMs with 1.58-bit weights substantially reduce model memory footprint, largely eliminating weight storage as the primary bottleneck. As context lengths grow, however, the bottleneck shifts to the key-value (KV) cache and attention, whose memory traffic increases linearly with sequence length and dominates long-context inference. Existing LLM accelerators mostly handle ternary linear layers and attention with separate datapaths, resulting in inefficient end-to-end execution and costly hardware overhead.
  To address these challenges, we propose \design, a sparse ternary LLM accelerator for efficient long-context edge inference. \design is built on three key techniques: (1) \design features a phase-aware sparse dataflow that jointly exploits heterogeneous sparsity across compute-bound prefill and memory-bound decode. (2) We design a unified compute fabric in \design to map ternary SpMM/SpMV and bit-serial attention onto a shared processing-element, efficiently exploiting both zero-weight and zero-bit sparsity. (3) To improve the efficiency of long-context inference, we propose an adaptive attention engine with two-stage selection. It first applies coarse-grained page-level Top-$k$ selection to reduce K-cache accesses, and then performs fine-grained token-level Top-$p$ pruning to further minimize V-cache traffic. 
  Across SOTA ternary LLMs, \design reduces KV traffic by 4.1--4.8$\times$ over full attention and 1.3--1.5$\times$ over a SOTA sparse-attention scheme with task-dependent quality loss. Against an NVIDIA Jetson AGX Orin GPU, \design achieves 3.9$\times$ higher generation throughput and 8.3$\times$ higher end-to-end energy efficiency on average; against an AMD Ryzen~7 9700X CPU, the improvements are 5.0$\times$ and 81.7$\times$, respectively. Against SOTA LLM accelerators, \design achieves 1.6--12.0$\times$ higher throughput than FIGLUT, TeLLMe, and MCBP, 2.0--4.2$\times$ higher energy efficiency, and 1.3--51.4$\times$ higher area efficiency.
\end{abstract}

\maketitle

\section{Introduction}
\label{sec:intro}

Large language models (LLMs) have achieved remarkable proficiency in open-ended dialogue, code generation, mathematical reasoning, and long-document understanding~\cite{touvron2023llama,chen2021evaluating}, with recent advances extending to multi-modal perception and generation. LLM deployment is increasingly migrating from the cloud to the edge, where personal assistants, in-vehicle copilots, and robots run inference locally to protect private data, avoid network latency, and eliminate token costs~\cite{gerganov2023llamacpp,team2025minicpm4}. However, edge devices impose stringent memory, bandwidth, and power constraints, which become particularly severe for long-context inference.

\begin{figure}[t]
  \centering
  \includegraphics[width=\linewidth]{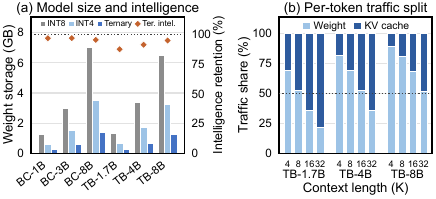}
  \caption{(a) Weight storage and intelligence retention vs.\ full-precision base models; (b) per-token weight and KV traffic.}
  \label{fig:motivation_model}
\end{figure}

Model weights and attention are the two dominant memory bottlenecks in long-context edge inference. Prior work reduces model size through low-bit quantization~\cite{frantar2022gptq,lin2024awq,xiao2023smoothquant,zhao2024atom,dettmers2023spqr}, while LUT-based designs accelerate mixed-precision execution~\cite{park2025figlut}. As shown in \figref{fig:motivation_model}(a), recent ternary LLMs~\cite{ma2024era,team2025minicpm4,prismml2026ternarybonsai,huang2026sherry} further compress weights to 1.58 bits using $\{-1,0,+1\}$ while preserving 88--97\% of full-precision intelligence with only one-fifth of the storage of 8-bit models. Their ternary arithmetic also replaces multiplication with add, subtract, or skip operations. However, as the context grows to tens of thousands of tokens, the KV cache scales linearly and dominates per-token memory traffic, as shown in \figref{fig:motivation_model}(b). Ternary quantization therefore alleviates weight traffic but exposes KV cache traffic as the new dominant bottleneck.

As summarized in Table~\ref{tab:soa-accelerators}, existing LLM accelerators~\cite{wang2025mcbp,li2026veclut,qiao2025tellme,huang2025tenet,guan2026tom,lin2026vitallm} optimize only subsets of the end-to-end inference pipeline and fall short in three key aspects:

(1) \textbf{Limited exploitation of heterogeneous sparsity across inference phases}: Ternary LLMs exhibit multiple forms of sparsity, including static zero-weight sparsity~\cite{guan2026tom}, bit-level sparsity in attention computation~\cite{wang2025mcbp}, and dynamic attention sparsity~\cite{lin2026twilight}. Meanwhile, prefill is dominated by compute-bound SpMM (sparse--dense matrix multiplication), whereas decode is dominated by memory-bound SpMV (sparse matrix--vector multiplication) and repeated accesses to weights and the KV cache. Existing ternary accelerators mainly optimize either ternary linear computation~\cite{li2026veclut,qiao2025tellme,wang2025bitnet} or sparse attention~\cite{huang2025tenet}, but fail to jointly exploit these complementary sparsities. This leads to suboptimal end-to-end inference efficiency.

(2) \textbf{\rev{Limited datapath unification for sparse ternary inference}}: 
Ternary LLMs exhibit substantial precision heterogeneity: linear layers use $\{-1,0,+1\}$ weights, whereas attention requires multi-bit computation to preserve model quality. Existing designs~\cite{qiao2025tellme,wang2025bitnet,huang2025tenet} typically employ lookup-table-based ternary matrix-multiplication engines. However, these LUT-based PEs are optimized for dense ternary computation and cannot efficiently support sparse SpMM and SpMV at the same time. As a result, they fail to exploit the abundant zero-weight sparsity in ternary LLMs and often require separate datapaths for linear layers and attention~\cite{chen2025bitmod,lin2026vitallm}.

(3) \textbf{Inefficient sparse attention for long-context inference}:
Prior accelerators support either dense~\cite{qiao2025tellme,wang2025bitnet}, fixed-pattern~\cite{huang2025tenet}, or Top-$k$~\cite{tang2024quest,wang2025mcbp,lin2026vitallm} attention, making them poorly suited to long contexts. Fixed-pattern~\cite{huang2025tenet,zaheer2020big} schemes cannot capture input-dependent attention sparsity, while fixed Top-$k$ selection~\cite{tang2024quest,wang2025mcbp,lin2026vitallm} imposes a rigid token budget. A small budget may discard important tokens, whereas a large budget introduces redundant KV-cache accesses. Consequently, existing schemes cannot adapt the selected KV set to the highly variable attention distributions across layers, heads, and inputs.

\begin{table}[t]
  \centering
  \scriptsize
  \caption{Comparison of LLM accelerators.}
  \label{tab:soa-accelerators}
  \setlength{\tabcolsep}{3.5pt}
  \renewcommand{\arraystretch}{1.1}
  \begin{tabular}{l | l l l l l}
      \hline
      \textbf{Design} & \textbf{Precision} & \textbf{Linear} & \textbf{Attention} & \textbf{Efficiency} & \textbf{Max Seq.} \\
      \hline
      MCBP~\cite{wang2025mcbp} & W8/A8 & Sparse & Sparse (Top-$k$) & Medium & 8K \\
      Vec-LUT~\cite{li2026veclut} & W1.58/A8 & Dense & Dense & Low & - \\
      TeLLMe~\cite{qiao2025tellme} & W1.58/A8 & Dense & Dense & Low & 1K \\
      TENET~\cite{huang2025tenet} & W1.58/A8 & Dense & Sparse (Static) & Medium & 1.5K \\
      \rev{VitaLLM~\cite{lin2026vitallm}} & \rev{W1.58/A8} & \rev{Dense} & \rev{Sparse (Top-$k$)} & \rev{Medium} & \rev{4K} \\
      \hline
      \textbf{\design} & \textbf{W1.58/A8} & \textbf{Sparse} & \textbf{Sparse (Adaptive)} & \textbf{High} & \textbf{32K} \\
      \hline
  \end{tabular}
\end{table}

To address these limitations, we present \design, a sparse ternary LLM accelerator for long-context edge inference. \design systematically exploits heterogeneous sparsity across inference phases and kernels. As shown in Table~\ref{tab:soa-accelerators}, \design is, to the best of our knowledge, the first accelerator to efficiently support sequence lengths beyond 8K.

We make the following contributions:
\begin{itemize}[leftmargin=*,topsep=0pt,itemsep=0pt,parsep=0pt,partopsep=0pt]
  \item \textbf{Sparse Dataflow for Ternary LLMs}: We identify complementary sparsity across ternary weights, attention bit patterns, and KV accesses. We propose a phase-aware sparse dataflow that jointly exploits these sparsity opportunities across inference phases, enabling efficient long-context inference under edge resource constraints.

  \item \textbf{Unified Datapath for Sparse Inference}: We formulate ternary SpMM/SpMV and bit-serial QK/PV as a common outer-product conditional-accumulation primitive. Then, we propose a shared compute fabric design with a unified datapath for various sparsity patterns. The proposed design can be reused to accelerate compute-bound prefill, memory-bound decode, ternary linear layers, and sparse attention, while benefiting from sparsity.
  
  \item \textbf{Adaptive Sparse Attention}: To improve long-sequence inference efficiency, we propose a two-stage adaptive attention scheme that combines coarse-grained page-level Top-$k$ selection with fine-grained token-level Top-$p$ pruning to progressively reduce K and V traffic. We address the implementation complexity with an adaptive sparse attention engine that integrates a bitonic Top-$k$ selector and a streaming-histogram Top-$p$ engine to efficiently realize the proposed dataflow.
  
  \item \textbf{Key Results}: Across SOTA ternary LLMs~\cite{prismml2026ternarybonsai,bitcpmcann2026}, \design reduces KV traffic by 4.1--4.8$\times$, with task-dependent quality loss. Compared with an NVIDIA Jetson AGX Orin GPU, \design achieves 3.9$\times$ higher generation throughput and 8.3$\times$ higher end-to-end energy efficiency; the corresponding gains over an AMD Ryzen~7 9700X CPU are 5.0$\times$ and 81.7$\times$. Against SOTA LLM accelerators~\cite{huang2025tenet,qiao2025tellme,park2025figlut,wang2025mcbp}, \design achieves 1.6--12.0$\times$ higher throughput than FIGLUT, TeLLMe, and MCBP, and 2.0--4.2$\times$ higher energy efficiency across prefill and decode.
\end{itemize}

%%%%%%%%%%%%%%%%%%%%%%%%%%%%%%%%%%%%%%%%%%%%%%%%%%%%%%%%%%%%%%%%%%%%%%%%%%%%%%%%%%%%%%%%%%
%%%%%%%%%%%%%%%%%%%%%%%%%%%%%%%%%%%%%%%%%%%%%%%%%%%%%%%%%%%%%%%%%%%%%%%%%%%%%%%%%%%%%%%%%%
\section{Background and Motivation}
\label{sec:bak_motivation}

\subsection{Ternary LLM Inference}
\begin{figure}[t]
  \centering
  \includegraphics[width=\linewidth]{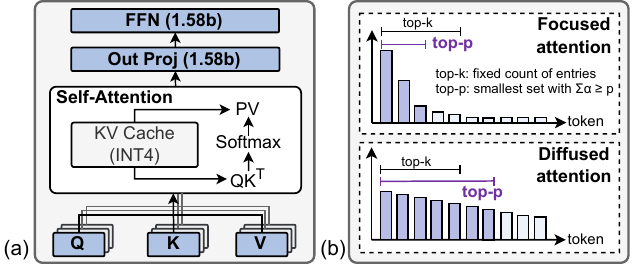}
  \caption{(a) Decoder block of ternary LLMs; (b) Content-aware sparse attention.}
  %% Centralized Attention
  \label{fig:ternary_llm_arch}
\end{figure}

Autoregressive inference has compute-bound \textit{prefill} and memory-bound \textit{decode}~\cite{pope2023efficiently}. Each decode token streams weights and accesses a sequence-growing KV cache. \figref{fig:ternary_llm_arch}(a) shows 1.58-bit ternary linear weights and INT4 KV cache for attention calculation.

\noindent\textbf{Linear Layer.}
Ternary constrains linear-layer weights to $\{-1,0,+1\}$ with FP16-comparable accuracy~\cite{ma2024era,prismml2026ternarybonsai}. It cuts FP16 weight traffic by roughly $10\times$ and replaces MACs with add, subtract, or skip. The sparsity in weights enables zero skipping, thereby enabling more efficient inference.

\noindent\textbf{Attention and KV-Cache.}
Though linear weights can be quantized to ternary values, the attention remains at higher precision to ensure model accuracy. QServe, KIVI, and KVQuant compress KV caches to about 4 bits~\cite{lin2024qserve,liu2024kivi,hooper2024kvquant}, while MCBP and PADE use INT8 $QK^{\mathsf T}$/$PV$ with FP16 nonlinear operations~\cite{wang2025mcbp,wang2026pade}. For a decode query $q$ over $N$ cached tokens, the dense attention is computed as follows:
\begin{equation}
  \alpha_i =
  \frac{\exp\left(q^{\mathsf T}k_i/\sqrt{D}\right)}
  {\sum_{j=1}^{N}\exp\left(q^{\mathsf T}k_j/\sqrt{D}\right)},
  \qquad
  o=\sum_{i=1}^{N}\alpha_i v_i.
  \label{eq:dense_attention}
\end{equation}

Eq.~\eqref{eq:dense_attention} scores every cached key and combines every value. Thus, KV-cache quantization reduces bytes per entry, but dense attention still need to access all KV data and incurs quadratic complexity.

\subsection{Sparse Attention}
Sparse attention instead reduces the number of KV entries accessed. Fixed-pattern methods retain query-independent sinks and local windows~\cite{xiao2024efficient,huang2025tenet}. They are hardware-friendly but may discard distant and relevant tokens, thus creating accuracy loss under long context.

To address this issue, content-aware Top-$k$ in \figref{fig:ternary_llm_arch}(b) selects a fixed-size set by exact or estimated relevance:
\begin{equation}
\begin{aligned}
\text{Top-}k:\quad
\mathcal I_k
&\in
\operatorname*{arg\,max}_{\substack{\mathcal I\subseteq\{1,\ldots,N\}\\
|\mathcal I|=k}}
\sum_{i\in\mathcal I}\tilde{s}_i,\\
\end{aligned}
\end{equation}
Here, $\tilde{s}_i$ is the relevance score and $\mathcal I_k$ is the retained set. Quest applies Top-$k$ at the KV-page level~\cite{tang2024quest}. Its fixed budget is predictable, but may over-fetch focused attention or discard useful context under diffused attention scores.

\begin{figure}[!t]
  \centering
  \includegraphics[width=\linewidth]{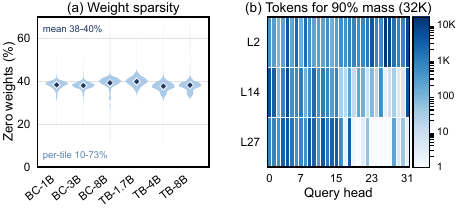}
  \caption{(a) Sparsity ratios of ternary LLMs; (b) median tokens covering 90\% attention mass per head (BitCPM-3B, 32K).}
  \label{fig:motivation_sparsity_mass}
\end{figure}

Top-$p$ is more adaptive to the distributions of attention scores. It removes the fixed-budget constraint by selecting the smallest set covering a target probability mass:
\begin{equation}
\begin{aligned}
\text{Top-}p:\quad
\mathcal I_p
&\in
\operatorname*{arg\,min}_{\mathcal I\subseteq\{1,\ldots,N\}}
|\mathcal I|
\quad\mathrm{s.t.}\quad
\sum_{i\in\mathcal I}\alpha_i\geq p.
\end{aligned}
\end{equation}
Here, $\alpha_i$ is the normalized probability from Eq.~\eqref{eq:dense_attention} and $\mathcal I_p$ is the retained set. Twilight combines coarse filtering with probability-mass-based Top-$p$ pruning~\cite{lin2026twilight}, adapting the retained count to each query. This adaptivity saves KV accesses but requires runtime selection and irregular data movement.

\subsection{Challenges and Opportunities}
\noindent\textbf{C1: Efficient Prefill and Decode Dataflows for Ternary LLMs.}
LLM inference exhibits \textbf{computational heterogeneity across inference phases}: prefill is dominated by compute-bound GEMM with high weight reuse, whereas decode is memory-bound due to GEMV and multi-head attention (MHA) that repeatedly stream model weights and the growing KV cache~\cite{patel2024splitwise,zhong2024distserve,agrawal2024taming}.
Although ternary LLMs greatly reduce weight storage, they introduce multiple sparsity opportunities, including zero-valued ternary weights, sparse bits in attention computation, and highly skewed attention distributions, as shown in \figref{fig:motivation_sparsity_mass}. Significant fractions of ternary weights and sparse attention values contribute no effective computation, while attention sparsity can eliminate up to 90\% of token accesses at long contexts.
However, existing accelerators mostly accelerate ternary linear layers and sparse attention separately~\cite{huang2025tenet,qiao2025tellme,guan2026tom,wang2021spatten,lu2021sanger,wang2024sofa}. They do not coordinate these complementary sparsities across prefill and decode, leaving end-to-end optimization opportunities unexploited.

\noindent\textbf{Solution:}
We propose a phase-aware sparse dataflow in \design that systematically exploits heterogeneous sparsity across prefill and decode. By jointly leveraging ternary-weight sparsity, bit-level sparsity, and adaptive attention sparsity, the proposed dataflow improves end-to-end efficiency for long-context ternary LLM inference.

\begin{figure}[!t]
  \centering
  \includegraphics[width=\linewidth]{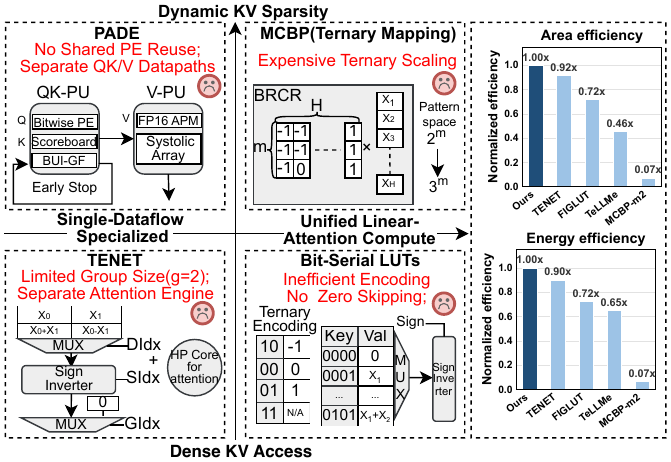}
  \caption{Design-space limitations of existing LLM accelerator PEs and
  normalized area and energy efficiency comparisons.}
  \label{fig:limit_pe}
\end{figure}

\begin{figure}[t]
  \centering
  \includegraphics[width=\linewidth]{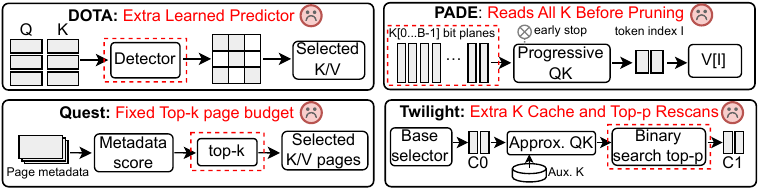}
  \caption{Tradeoffs of prior sparse-attention paths: an extra learned predictor, key scoring over the full context, a fixed Top-$k$ page budget, or an extra K cache with repeated Top-$p$ scans.}
  \label{fig:challenge_attn}
\end{figure}

\noindent\textbf{C2: Unified Datapath and Efficient PE Design.}
Ternary LLM inference introduces \textbf{operator-level precision and datapath heterogeneity}: linear layers use ternary weights, whereas attention uses multi-bit operands with sparse Q/P bit planes. \figref{fig:limit_pe} summarizes how existing PEs target this heterogeneity. \rev{PADE fuses predictor-free sparsity discovery with attention execution, leaving projections and FFNs to the GPU~\cite{wang2026pade}. TENET/TeLLMe use ternary LUTs~\cite{huang2025tenet,qiao2025tellme}, while FIGLUT/MCBP use binary-plane or base-3 mappings~\cite{park2025figlut,wang2025mcbp}. These designs do not share one datapath across ternary GEMM/GEMV and multi-bit QK/PV.}

\noindent\textbf{Solution:}
\design maps ternary GEMM/GEMV and bit-serial QK/PV to one conditional-accumulation datapath. It skips zero weights and zero Q/P bits, while bank-local accumulation and conflict FIFOs handle irregular accesses without LUT or pattern reconstruction. Compared with TENET and TeLLMe, the PE design in \design achieves up to 2.2$\times$ higher area efficiency and 1.5$\times$ higher energy efficiency, as shown in \figref{fig:limit_pe}.
\rev{Adapting MCBP's binary patterns to ternary weights via positive/negative passes reduces area but increases cycles; area and estimated energy efficiencies are only 7.3\% and 6.8\% of \design's.}

\noindent\textbf{C3: Adaptive Sparse Attention for Long-Context Inference.}
As context lengths grow, KV-cache accesses dominate decode traffic (\figref{fig:motivation_model}(b)), making sparse attention critical. However, fixed Top-$k$ selection cannot accommodate the highly variable attention sparsity across layers, heads, and inputs. As shown in \figref{fig:motivation_sparsity_mass}(b), the number of tokens covering 90\% of attention mass varies by three orders of magnitude, causing fixed budgets to either waste KV accesses or lose important tokens. Top-$p$ selection addresses this limitation by dynamically adjusting the retained tokens based on each attention distribution~\cite{holtzman2019curious,lin2026twilight}. 
Efficient adaptation, however, must avoid offsetting its traffic savings with selection overhead. \rev{As summarized in \figref{fig:challenge_attn}, prior designs make different tradeoffs: DOTA trains a lightweight predictor, PADE avoids a predictor by scoring keys bit-serially with early termination, Quest bounds K traffic with a fixed page budget, and Twilight estimates Top-$p$ from an extra low-precision K cache~\cite{qu2022dota,wang2026pade,tang2024quest,lin2026twilight}.}

\noindent\textbf{Solution:}
We propose an adaptive sparse attention with two-stage selection for efficient long-context attention. The proposed attention scheme is composed of a page-level Top-$k$ selection and a histogram-based Top-$p$ selection to adaptively prune KV traffic. We also present the low-complexity architecture to implement the adaptive sparse attention on area-constrained hardware.

%%%%%%%%%%%%%%%%%%%%%%%%%%%%%%%%%%%%%%%%%%%%%%%%%%%%%%%%%%%%%%%%%%%%%%%%%%%%%%%%%%%%%%%%%%
%%%%%%%%%%%%%%%%%%%%%%%%%%%%%%%%%%%%%%%%%%%%%%%%%%%%%%%%%%%%%%%%%%%%%%%%%%%%%%%%%%%%%%%%%%
\section{\design Dataflow} \label{sec:dataflow}
%% GEMM/GEMV/bit-serial数据流映射

\subsection{Phase-aware Sparse Dataflow}
\label{sec:dataflow_flow}
% Chunk Prefill, Attention, 精度分配

\begin{figure}[!t]
  \centering
  \includegraphics[width=\linewidth]{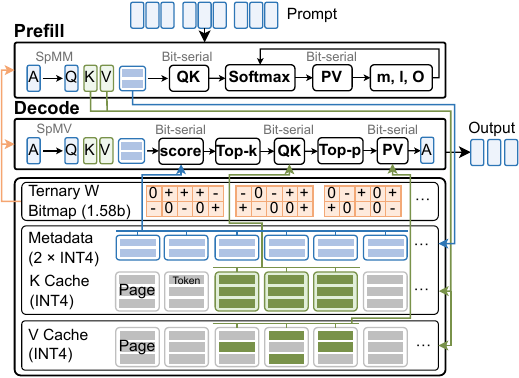}
  \caption{Overview of the prefill and decode dataflows.}
  \label{fig:dataflow_overview}
\end{figure}

%% 精简 + Ternary 的特殊性（1. 权重稀疏）+ Attention
%% 不要说 SpMM/SpMV 这些，这些放到下两节细说
Prefill exposes token parallelism and weight reuse, whereas single-token decode repeatedly streams model weights and the growing KV cache~\cite{pope2023efficiently}. Supporting both phases requires exploiting zero ternary weights and sparse Q/P bits in computation, while converting query-dependent attention sparsity into fewer K/V accesses~\cite{guan2026tom,wang2025mcbp,lin2026twilight}. Coordinating these opportunities without phase- or operator-specific paths is the central dataflow challenge.
\figref{fig:dataflow_overview} illustrates how \design integrates these mechanisms across prefill and decode. The unified sparse execution in Section~\ref{sec:dataflow_sparse} maps linear and attention computation onto a unified execution path, while the adaptive attention with two-stage selection in Section~\ref{sec:dataflow_selection} progressively reduces KV cache traffic.

%% Prefill/Decode 加粗
\noindent\textbf{Prefill.}
The prompt is partitioned into fixed-size chunks to bound the on-chip working set. Within each Transformer layer, bitmap-compressed ternary weights are reused across the tokens in each chunk for the QKV projections, output projection, and FFN. The generated K/V vectors are quantized to INT4 and appended to the paged cache. Each completed K page also records a signed INT4 minimum and maximum per feature as compact metadata for decode.

Prefill attention scans the cached K/V chunks using online Softmax~\cite{milakov2018online,dao2022flashattention}. Only the running maximum, normalizer, and output accumulator $(m,\ell,\mathbf{O})$ persist across chunks, bounding the on-chip attention state while producing the exact result.

\noindent\textbf{Decode.}
After the prompt is consumed, each generated token triggers one decode pass through the model. The linear layers stream weights for the new activation vector. Attention first selects pages from metadata, computes exact candidate QK scores for tokens in those pages, prunes the resulting tokens by probability mass, and finally fetches the retained V data for PV.

\subsection{Unified Sparse Execution for Linear and Attention} \label{sec:dataflow_sparse}
% \zyz{examples: One significant limitation of prior ternary LLM accelerators~\cite{} is that they fail to or partly exploit the weight or activation sparsity. This degrades the inference efficiency because quite amount of zero-related operations do not contribute to the computation results. We observe that ternary LLMs exhibit several levels of sparsity: (1) weight sparsity and (2) activation sparsity. }
Ternary linear layers and multi-bit attention expose sparsity in different operand representations. Linear layers contain static zero weights, whereas bit-serial attention contains zero bits across the Q and P bit planes; both eliminate vector updates that do not contribute to the result. Existing accelerators typically specialize in one class of operations or use separate datapaths for the two~\cite{park2025figlut,qiao2025tellme,huang2025tenet,wang2026pade}, limiting their ability to exploit both forms of sparsity within a unified design. Supporting both on one compute structure is difficult because ternary SpMM/SpMV encodes each update with a ternary weight, while QK/PV decomposes multi-bit Q/P coefficients into shifted bit-plane updates and maps parallel lanes to different dimensions.

\begin{figure}[t]
  \centering
  \includegraphics[width=\linewidth]{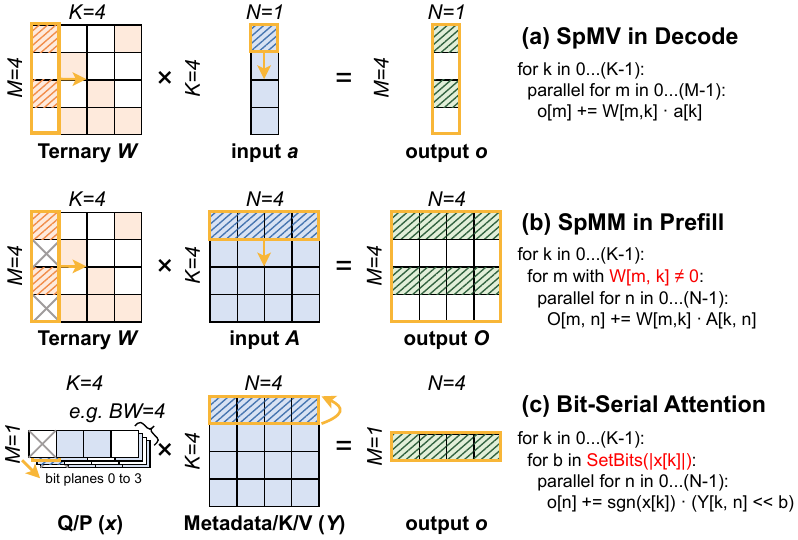}
  \caption{Unified dataflows for ternary SpMM/SpMV and bit-serial attention.}
  \label{fig:unified_compute_model}
\end{figure}

For the shared dataflow, neither the dot-product nor the row-wise organization is well suited to the compute structure. In a dot-product dataflow, each lane computes one output, so varying nonzero counts make lanes finish at different times, and every output requires a separate reduction. A row-wise dataflow completes one output row at a time, but sparse column indices cause irregular accesses to dense-matrix rows and limit their reuse. \design therefore adopts an outer-product dataflow, which provides regular vector updates and reuses the selected operand across multiple lanes.

\figref{fig:unified_compute_model} shows how \design maps all three operators to the outer-product dataflow. At each outer-product step, a selected vector operand $\mathbf{v}$ is optionally scaled by a power of two and added to or subtracted from an accumulator state $\mathbf{z}$:
\begin{equation}
\mathbf{z} \leftarrow \mathbf{z} + s\,2^b\mathbf{v}, \qquad s\in\{-1,+1\}.
\label{eq:conditional_accumulation}
\end{equation}
In ternary SpMM, each nonzero weight updates a vector of output columns, while SpMV uses one activation to update multiple output rows; both derive $s$ from the ternary weight and use $b=0$. Bit-serial attention decomposes each multi-bit coefficient into a sequence of shifted vector updates. In \figref{fig:unified_compute_model}(c), $x_{b,k}\in\{0,1\}$ denotes bit $b$ of the magnitude $|x_k|$: $x$ represents Q during page and QK scoring, and P during PV; $Y$ is the corresponding metadata, K, or V operand. Metadata scoring and QK derive $s$ from the query sign and $b$ from its set-bit position. PV represents each normalized probability with 16 fixed-point probability bits (P-bits); every set-bit position supplies $b$, with $s=+1$. Although the vector operands and lane meanings differ, every operator is expressed as the same conditional vector update. This common execution model removes the need for operator-specific accumulation datapaths; Section~\ref{sec:arch_fabric} details the hardware and lane mappings.

\subsection{Adaptive Sparse Attention with Two-Stage Selection} \label{sec:dataflow_selection}
% \zyz{briefly analyze the key challenges of existing sparse attention strategies. then explain why our proposed sparse attention scheme is needed. finally dive into our design. for example: The data movement due to KV cache is a major challenge for long-context decoding as shown in \figref{fig:motivation_model}. Existing sparse attention schemes adopt fixed pattern~\cite{} or top-$k$~\cite{} strategies to bypass/filter unimportant KV tokens, thereby improving the decoding throughput. However, these schemes fails to xxx and is still costly for long-context scenarios because xxxxxx. To resolve this issue, we propose a memory-efficient adaptive sparse attention with two-stage selection.}

%% 别人 top-k 的公式
%% 我们的 top-k+top-p 公式
The goal of adaptive sparse attention is to reduce KV traffic with low selection overhead. Top-$k$ bounds K traffic but cannot adapt its fixed budget to different attention score distributions~\cite{tang2024quest,yang2024post}. Top-$p$ adapts to the candidate attention distribution but requires exact QK scores and thus cannot reduce K reads~\cite{lin2026twilight}. Combining them allows Top-$k$ to prune K pages first and Top-$p$ to further prune V tokens.

Let $N$ denote the total number of cached tokens for the current query, and let $\rho_k$ be the fraction retained by page-level Top-$k$. The selected pages form a candidate set $\mathcal{T}$ with $|\mathcal{T}|=\rho_kN$ tokens, including the mandatory sink and recent pages. Top-$p$ further retains a set $\mathcal{R}\subseteq\mathcal{T}$, where $\rho_p=|\mathcal{R}|/(\rho_kN)$ is the retained fraction among the candidates. \rev{\key{The page budget is $k=\lceil\rho_k\lceil N/16\rceil\rceil$, e.g., 512 pages for a 32K-token context at $\rho_k=1/4$, the largest configuration the selector supports.}}

Twilight demonstrates the effectiveness of hierarchical Top-$k$+Top-$p$ pruning~\cite{lin2026twilight} on GPU. Pruner requires an additional low-precision K scoring pass. \design redesigns this path for native paged-INT4 execution: compact metadata selects K pages, and the resulting candidate QK scores directly drive Top-$p$. To quantify the resulting traffic savings, we count one token's KV vectors as one traffic unit. The traffic $C$ under dense, Top-$k$-only, and Top-$k$+Top-$p$ attention is:
\begin{equation}
\begin{aligned}
C_{\mathrm{dense}} &= 2N,\\
C_{k} &= \frac{N}{8}+2\rho_kN,\\
C_{k+p} &= \frac{N}{8}+(1+\rho_p)\rho_kN.
\end{aligned}
\label{eq:kv_traffic_model}
\end{equation}
Here, the $N/8$ term reads two metadata extrema per channel for every 16-token page. Top-$k$-only attention reads both K and V for all $\rho_kN$ candidates. Adding Top-$p$ leaves the K traffic unchanged but reduces V traffic to $\rho_p\rho_kN$.

These traffic savings translate into speedup only if Top-$p$ selection is itself hardware-efficient. The main challenge is finding the mass threshold over $\rho_kN$ candidate scores. Twilight\cite{lin2026twilight} uses binary search and rescans the candidates $L$ times, requiring $O(\rho_kNL)$ work; its stopping tolerance of $10^{-5}$ allows up to $L=17$ full scans of normalized attention weights. In contrast, \design streams the candidate masses into a fixed $B$-bin histogram and scans only the bins for the threshold, reducing the work to $O(\rho_kN+B)$ with fixed-size on-chip state.
Algorithm~\ref{alg:two_stage_selection} summarizes the resulting decode flow.

% \zyz{add how we estimate kv traffic here using equations}

\begin{algorithm}[t]
\caption{Two-stage Top-$k$ + Top-$p$ decode attention}
\label{alg:two_stage_selection}
\small
\begin{algorithmic}[1]
\REQUIRE Query $\mathbf{q}$; page metadata; page budget $k$; mass target $p$
\ENSURE Sparse attention output $\mathbf{o}$
\STATE Score all pages using Eq.~\eqref{eq:page_proxy_score}
\STATE Select $\mathcal{G}$ within the total budget $k$; form $\mathcal{T}$; gather only $K[\mathcal{T}]$
\STATE Compute exact candidate QK scores for $t\in\mathcal{T}$ and find $m$
\STATE Quantize each $\Delta_t$; accumulate $H$ and record $b_t$
\STATE Scan $H$ to obtain $b^\star$ and retained tokens $\mathcal{R}$
\STATE Gather only $V[\mathcal{R}]$
\STATE Compute $\mathbf{o}$ using Eq.~\eqref{eq:sparse_pv_output}
\end{algorithmic}
\end{algorithm}

\noindent\textbf{K Page Selection with Top-$k$.} Let $K^{\min}_{g,d}$ and $K^{\max}_{g,d}$ be the signed INT4 extrema of key dimension $d$ in page $g$. For query $\mathbf{q}$, the page score and selected page set are
\begin{equation}
\begin{aligned}
\hat{s}_g &= \sum_{d=0}^{D-1}\max\left(q_dK^{\min}_{g,d},\,q_dK^{\max}_{g,d}\right),\\
\mathcal{G} &= \mathcal{G}_{\mathrm{keep}}\cup
\operatorname{TopK}_{k-|\mathcal{G}_{\mathrm{keep}}|}\!
\left(\{\hat{s}_g:g\notin\mathcal{G}_{\mathrm{keep}}\}\right),
\end{aligned}
\label{eq:page_proxy_score}
\end{equation}
where $\mathcal{G}_{\mathrm{keep}}$ contains the mandatory sink and recent pages, which count toward the total budget $k$. For $k\geq|\mathcal{G}_{\mathrm{keep}}|$, it gives $|\mathcal{G}|=k$. Because each key element lies between its page extrema, $\mathbf{q}^{\mathsf T}\mathbf{k}_{g,t}\leq\hat{s}_g$ for every token $t$ in page $g$. Thus, page selection only reads metadata and K vectors are fetched only after their pages enter $\mathcal{G}$, forming the candidate set $\mathcal{T}=\bigcup_{g\in\mathcal{G}}\operatorname{tokens}(g)$ with $|\mathcal{T}|=\rho_kN$.

\noindent\textbf{V Token Selection with Top-$p$.} After gathering $K[\mathcal{T}]$, \design computes the exact candidate QK score $s_t=\mathbf{q}^{\mathsf T}\mathbf{k}_t/\sqrt{D}$ of every candidate token. It then quantizes their exponential masses into the ordered histogram used for threshold search. Let $m=\max_{t\in\mathcal{T}}s_t$ be the largest candidate score and $\Delta_t=m-s_t\geq0$ be token $t$'s gap from that maximum. A smaller gap indicates a larger attention weight. The quantizer maps this gap to a histogram bin $b_t=Q(\Delta_t)$, with bin 0 closest to the maximum, and the exponential LUT returns the token's approximate unnormalized Softmax mass $\mu_t=\operatorname{ExpLUT}(b_t)\approx e^{-\Delta_t}$. The mass histogram and crossing bin are
\begin{equation}
\begin{aligned}
H_b &= \sum_{t\in\mathcal{T}:\,b_t=b}\mu_t,\\
b^\star &= \min\left\{b:\sum_{j=0}^{b}H_j\geq p\sum_{j}H_j\right\},\\
\mathcal{R} &= \{t\in\mathcal{T}:b_t\leq b^\star\}.
\end{aligned}
\label{eq:mass_histogram_selection}
\end{equation}
Here, $H_b$ is the total candidate mass in bin $b$. The bins are scanned from high scores to low scores, starting at bin 0, and their masses are accumulated until the running sum reaches or exceeds $p$ times the total mass within $\mathcal{T}$. The bin where this crossing occurs is $b^\star$. The retained set $\mathcal{R}$ includes every token from bin 0 through $b^\star$, conservatively meeting the candidate-set histogram mass target. \rev{\key{The target $p$ is normalized over $\mathcal{T}$ only, and Top-$p$ never revisits pages excluded by Top-$k$.}} Only $V[\mathcal{R}]$ is gathered, and PV renormalizes over the retained mass:
\begin{equation}
\mathbf{o}=\frac{\sum_{t\in\mathcal{R}}\mu_t\mathbf{v}_t}{\sum_{t\in\mathcal{R}}\mu_t}.
\label{eq:sparse_pv_output}
\end{equation}
\begin{figure}[!t]
  \centering
  \includegraphics[width=\linewidth]{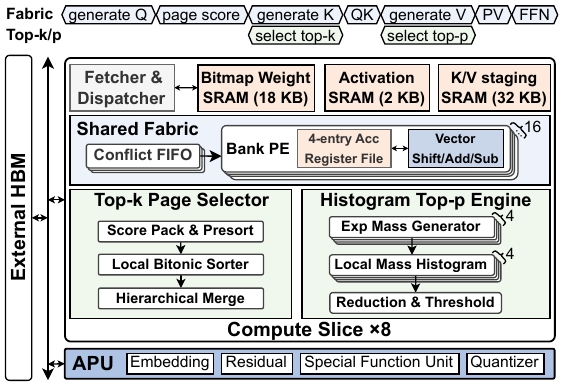}
  \caption{Architecture overview of \design.}
  \label{fig:arch_overview}
\end{figure}
\noindent\rev{\textbf{Grouped-Query Attention.} All evaluated models use GQA, where $G$ query heads share one KV head ($G=2$--$16$). Independent per-head selection would multiply the K traffic of a group by up to $G$, \key{so \design performs one page Top-$k$ per KV group}: each head converts its page scores $\hat{s}_g$ into a probability distribution over pages with a Softmax, and the group keeps the $k$ pages with the largest total probability across its heads. The heads share $\mathcal{T}$ and its K reads, Top-$p$ remains per head, and V is gathered for the union of the retained sets. \key{MQA is the single-group case.}}

%%%%%%%%%%%%%%%%%%%%%%%%%%%%%%%%%%%%%%%%%%%%%%%%%%%%%%%%%%%%%%%%%%%%%%%%%%%%%%%%%%%%%%%%%%
%%%%%%%%%%%%%%%%%%%%%%%%%%%%%%%%%%%%%%%%%%%%%%%%%%%%%%%%%%%%%%%%%%%%%%%%%%%%%%%%%%%%%%%%%%
\section{\design Architecture and Hardware}
\label{sec:arch}

\subsection{Architecture Overview}
\label{sec:arch_overview}

\design implements the aforementioned dataflow mainly through two key hardware components. First, a shared compute fabric implements the unified sparse kernels (SpMM, SpMV, and sparse attention), eliminating separate operator-specific modules. Second, it includes an adaptive sparse attention engine with a fixed-size Bitonic Top-$k$ page selector and a streaming histogram Top-$p$ engine. 

\figref{fig:arch_overview} shows the top-level architecture of \design, consisting of eight identical compute slices connected to external HBM2. The slices operate in parallel on independent attention-head and output-tile partitions. Each slice contains the shared fabric (Section~\ref{sec:arch_fabric}), the adaptive sparse attention engine (Section~\ref{sec:arch_adap_attn_engine}), operand buffers, and an activation-processing unit (APU). A fetcher and dispatcher stage three operand streams from HBM into per-slice SRAM: bitmap-compressed ternary weights, INT8 activations/queries, and packed-INT4 K/V data with their page extrema. The APU handles embedding, residual, special-function, and quantization operations.

The timing diagram at the top of \figref{fig:arch_overview} highlights inter-stage parallelism during decode. Because sink and recent pages are always retained (Section~\ref{sec:dataflow_selection}), \rev{K and V generation for the new token proceed in parallel with page Top-$k$ and histogram Top-$p$ selection, respectively.} Section~\ref{sec:eval_hardware} quantifies these overlaps.

\begin{figure}[!t]
  \centering
  \includegraphics[width=\linewidth]{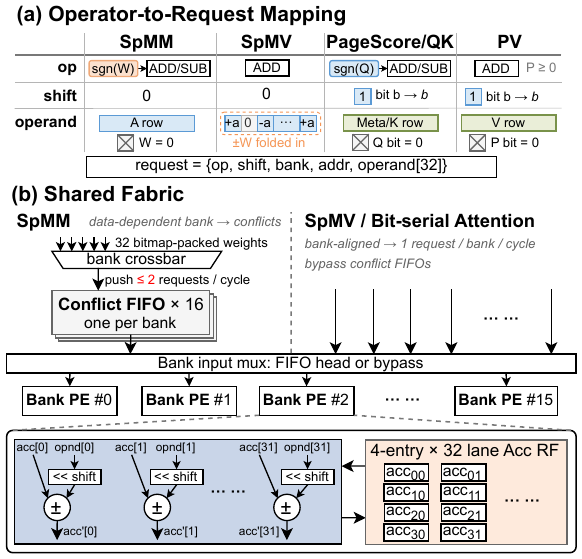}
  \caption{Operator-to-request mapping and the shared fabric.}
  \label{fig:shared_fabric}
\end{figure}

\begin{figure*}[t]
  \centering
  \includegraphics[width=\linewidth]{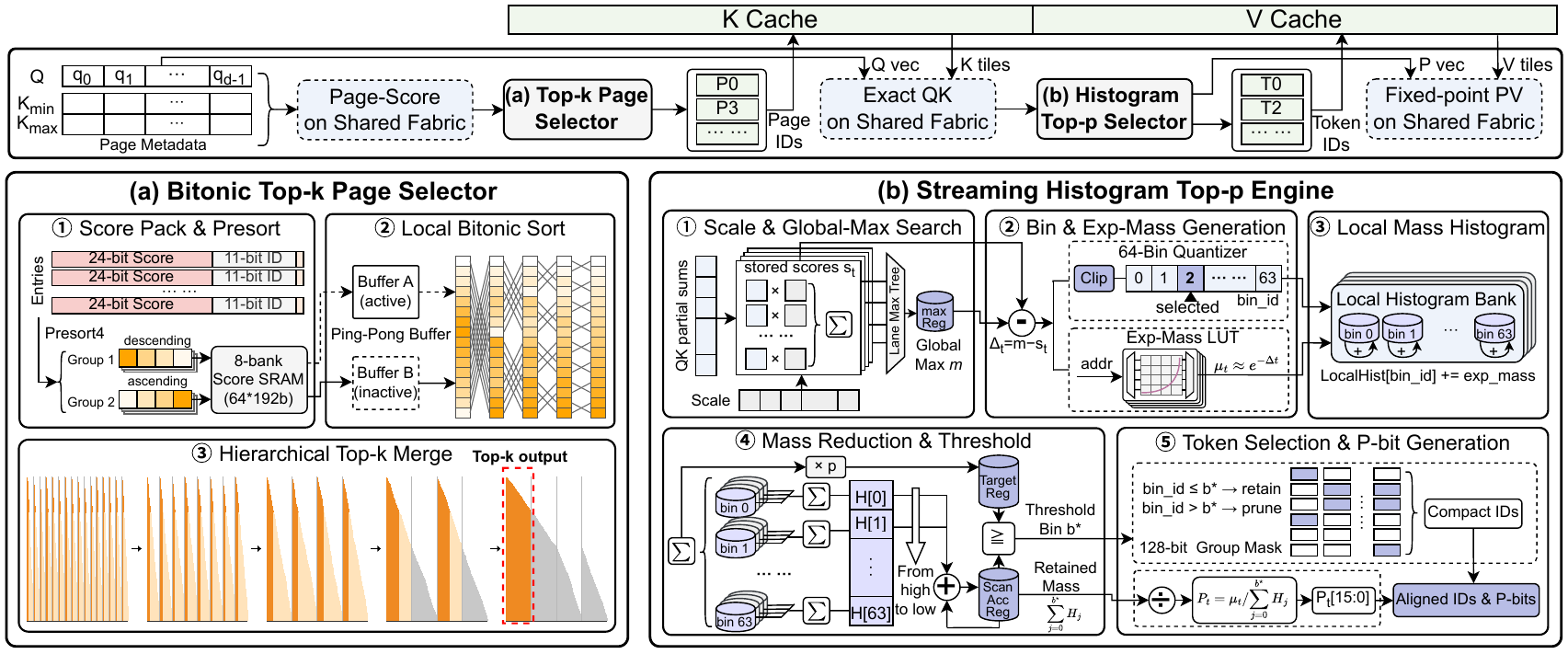}
  \caption{Two-stage sparse-attention selectors: (a) Bitonic page Top-$k$; (b) histogram token Top-$p$.}
  \label{fig:topk_topp_selector}
\end{figure*}

\subsection{Shared Fabric for Unified Sparse Execution}
\label{sec:arch_fabric}

SpMM, SpMV, and attention share the conditional update in Eq.~\eqref{eq:conditional_accumulation}, but expose sparsity along different dimensions. To preserve these sparsity opportunities on one fabric, \design represents each effective vector update with a common request containing an opcode, a shift amount, a target bank and accumulator address, and an operand vector, as shown in \figref{fig:shared_fabric}(a). The controller selects one request stream to drive the fabric until the operation completes. Switching between operations therefore changes only the request interpretation, while the PE array and output path remain unchanged.

\noindent\textbf{Ternary SpMM.}
Each nonzero weight generates an ADD or SUB request according to its sign, with zero shift and an activation row as the operand; the lanes map to output columns. Let $s_w$ denote the fraction of zero weights (about 40\%), and let $P$ and $L$ be the numbers of PEs and lanes per PE in one compute slice, respectively. Emitting requests only for nonzero weights reduces the ideal SpMM cycle complexity from $\mathcal{O}(MKN/(PL))$ to $\mathcal{O}((1-s_w)MKN/(PL))$, yielding an ideal speedup of $1/(1-s_w)\approx1.67\times$.

\noindent\textbf{Ternary SpMV.}
For an activation scalar $a$, a SpMV request uses ADD with zero shift, while its operand vector folds each ternary weight into the broadcast activation as $+a$, $0$, or $-a$; the lanes map to output rows. SpMV retains its bank-aligned group schedule, so its ideal cycle complexity remains $\mathcal{O}(MK/(PL))$, but gating zero-weight lanes reduces the useful arithmetic work from $\mathcal{O}(MK)$ to $\mathcal{O}((1-s_w)MK)$.

\noindent\textbf{Bit-serial Attention.}
Page scoring and QK encode the query sign as ADD/SUB and each set Q bit as a shift, with metadata or K rows as operands. PV uses ADD with each set P bit as a shift and V rows as operands. Let $s_q$ and $s_p$ be the zero-bit fractions of the $B_q$-bit queries and $B_p$-bit probabilities, and let $D$ denote the attention-head dimension. Issuing requests only for set bits reduces the ideal QK and PV cycle complexities to $\mathcal{O}((1-s_q)B_q\rho_kND/(PL))$ and $\mathcal{O}((1-s_p)B_p\lvert\mathcal{R}\rvert D/(PL))$; with $s_q\approx69\%$ and $s_p\approx83\%$, this yields $3.2\times$ and $5.7\times$ speedups, respectively.

As shown in \figref{fig:shared_fabric}(b), skipping zero weights makes the SpMM request stream irregular. Because each surviving weight determines its own destination, multiple requests may target the same bank in one cycle. A bank crossbar routes the requests by destination, and each bank uses a conflict FIFO that can accept up to 2 requests per cycle. The two enqueue ports allow the bitmap decoder to continue issuing requests while each FIFO drains one request per cycle. SpMV and bit-serial attention retain bank-aligned request streams with at most one request per bank per cycle; they therefore bypass the conflict FIFOs and enter the bank input mux directly.

The PE datapath in \figref{fig:shared_fabric}(b) uses $P=16$ PEs and $L=32$ accumulation lanes per PE, together with a 4-entry accumulator register file. Each request identifies a target PE and accumulator address. The corresponding partial sum remains in the local accumulator register file and is emitted directly when complete. Section~\ref{sec:eval_design_params} evaluates the choices of $P$, $L$, and the conflict-FIFO depth.

\subsection{Adaptive Sparse Attention Engine}\label{sec:arch_adap_attn_engine}

\noindent \textbf{Dataflow.}
The upper lane of \figref{fig:topk_topp_selector} summarizes the complete sparse-attention dataflow. The shared fabric first computes metadata-based page scores, and the Top-$k$ selector returns page IDs for K gathering. Exact QK scores over the gathered K tiles then feed the Top-$p$ engine, which selects token IDs, generates P-bits, and triggers V gathering. Finally, the resulting P-bits and gathered V tiles drive PV on the shared fabric.

\noindent \textbf{Bitonic Top-$k$ Page Selector.}
%% 对 cycle 的影响，性能
Page-level Top-$k$ must rank a page-score stream that grows with context length. As shown in \figref{fig:topk_topp_selector}(a), \design time-multiplexes a fixed-size Bitonic sorter, so a longer context increases the number of iterations rather than the sorter size.

Selection proceeds in three stages. Score Pack \& Presort (\circled{1}) pairs each score with its page ID, presorts every four entries, and writes them to the score SRAM. Local Bitonic Sort (\circled{2}) uses two ping-pong buffers and a small sorting network to process one window at a time. Finally, Hierarchical Top-$k$ Merge (\circled{3}) combines the sorted windows. 
\rev{All page scores are buffered in a 2048-entry page-score SRAM, so no rate matching is needed. Bitonic merges grow sorted runs up to the 512-entry selector capacity, reading one run in reverse (the descending window in \figref{fig:topk_topp_selector}(a)); beyond that, each merge keeps only the top half of its output.}

The sorter is reused across all windows and merge levels. Non-power-of-two page counts are padded with invalid entries that always rank below valid pages. Entries are ordered by score, with the smaller page ID selected when two scores are equal. Each gathered K tile is staged locally and reused across the set Q bits; unselected pages generate no HBM requests.

\noindent \textbf{Streaming Histogram Top-$p$ Engine.}
%% 如果实现了 binary searching，要遍历多少遍。
Top-$p$ must find a variable-size token set whose cumulative probability reaches the target $p$. Software usually finds the threshold by sorting or repeatedly scanning all candidate scores, which is costly in hardware. \design instead quantizes the exponential masses into a fixed-resolution histogram and finds the threshold with two fixed passes over 64 bins. \figref{fig:topk_topp_selector}(b) shows the resulting five-step engine.

As partial sums for the exact candidate QK scores arrive, Scale \& Global-Max Search (\circled{1}) applies the quantization scales, stores each final score $s_t$, and reduces the lane maxima to the global maximum $m$. Because each bin ID is defined relative to $m$, the final histogram cannot be updated before the global maximum is known. Bin \& Exp-Mass Generation (\circled{2}) therefore rereads the stored scores, quantizes each gap $m-s_t$ into one of 64 ordered bins, and uses the exponential-mass LUT to obtain $\mu_t\approx e^{s_t-m}$. Four parallel mass-generation lanes feed four local histogram banks, avoiding simultaneous updates to one histogram SRAM. Local Mass Histogram (\circled{3}) accumulates each $\mu_t$ into its local bin. \rev{The histogram, LUT, and reducer are fixed-size, and the candidate-score buffer is sized for the 8192-candidate maximum; only the latency of scaling and binning the candidate scores (steps \circled{1} and \circled{2}) grows with the number of candidates.}

Mass Reduction \& Threshold (\circled{4}) reduces the per-bank histograms into the global bin masses $H_b$. One bin pass computes the total mass and the target $p\sum_b H_b$. A second pass scans from bin 0 toward lower scores until the cumulative mass reaches the target, producing $b^\star$ and the retained mass $\sum_{j=0}^{b^\star}H_j$. Token Selection \& P-bit Generation (\circled{5}) applies $b_t\leq b^\star$ and compacts the survivors into aligned token-ID/P-bit pairs, with the P-bits encoding their normalized masses. A lookahead gather prefetches only $V[\mathcal{R}]$ and buffers the returned tiles until PV consumes them; the V of a token is skipped only when no head in its KV group retains it.

Section~\ref{sec:eval_design_params} evaluates both design tradeoffs: the Top-$k$ window size balances latency and sorter area, while the Top-$p$ histogram resolution balances threshold-search latency, SRAM footprint, and implementation complexity.

%%%%%%%%%%%%%%%%%%%%%%%%%%%%%%%%%%%%%%%%%%%%%%%%%%%%%%%%%%%%%%%%%%%%%%%%%%%%%%%%%%%%%%%%%%
%%%%%%%%%%%%%%%%%%%%%%%%%%%%%%%%%%%%%%%%%%%%%%%%%%%%%%%%%%%%%%%%%%%%%%%%%%%%%%%%%%%%%%%%%%
\section{Evaluation}
\label{sec:eval}

\subsection{Experimental Setup}

\noindent\textbf{Models and Workload.}
We evaluate six ternary LLMs from two model families: BitCPM-1B/3B/8B~\cite{bitcpmcann2026} and Ternary-Bonsai-1.7B/4B/8B~\cite{prismml2026ternarybonsai}. Algorithmic quality is measured by perplexity on WikiText-2~\cite{merity2017pointer} and PG19~\cite{rae2020compressive} and by task scores on seven generative benchmarks. GovReport~\cite{huang2021efficient}, QMSum~\cite{zhong2021qmsum}, and MultiNews~\cite{fabbri2019multinews} cover long-context summarization, while GSM8K~\cite{cobbe2021training}, MBPP~\cite{austin2021program}, IFEval~\cite{zhou2023ifeval}, and CoQA~\cite{reddy2019coqa} cover short-context reasoning, code generation, instruction following, and question answering.

\noindent\rev{\textbf{Quantization.} All quantizers are symmetric without zero-points: activations and queries use per-token INT8 scales, P uses 16-bit fixed point, and the KV cache uses INT4. BitCPM uses per-token scales for both K and V, while Bonsai follows KIVI~\cite{liu2024kivi} with per-channel K scales and per-token V scales.}

\noindent\textbf{Algorithm Baselines.} Model accuracy evaluations follow the settings in Twilight~\cite{lin2026twilight}. The dense baseline performs full attention under the same numerical formats, while Quest~\cite{tang2024quest} is used as the page-level Top-$k$ baseline.

\begin{table}[!t]
  \centering
  \footnotesize
  \caption{Hardware configuration of \design.}
  \label{tab:hardware_config}
  \setlength{\tabcolsep}{2.5pt}
  \renewcommand{\arraystretch}{1.18}
  \begin{tabular}{>{\raggedright\arraybackslash}m{0.28\columnwidth}|p{0.65\columnwidth}}
      \specialrule{0.9pt}{0pt}{0pt}
      \textbf{Component} & \textbf{Configuration} \\ \hline
      Technology and Freq. & 28~nm, 1~GHz \\ \hline
      Organization & 8 Compute Slices; 16 PEs/Slice; 32 Lanes/PE \\ \hline
      Selection Engines & 64-entry Top-$k$ window; 64-bin Top-$p$ histogram \\ \hline
      On-chip Buffer & 144~KiB weights; 16~KiB Q tiles; 280~KiB K/V staging; 384~KiB QK scores; 256~KiB mass/scale/P-bit state; 16.25~KiB selection state \\ \hline
      External Memory & HBM2, $8{\times}128$-bit HBM channels @2~GHz, 8~GB \\
      \specialrule{0.9pt}{0pt}{0pt}
  \end{tabular}

\end{table}

\noindent\textbf{Hardware Baselines.} 
We compare the end-to-end inference performance of \design against recent edge LLM accelerators on CPU, GPU, FPGA, and ASIC platforms as follows:
% \zyz{revise}
\begin{itemize}[leftmargin=1.25em,itemsep=1pt,topsep=2pt,parsep=0pt]
\item \textbf{CPU and GPU.} We evaluate an AMD Ryzen~7 9700X CPU using the \texttt{llama.cpp}~\cite{gerganov2023llamacpp} SIMD backend with eight threads and an NVIDIA Jetson AGX Orin configured at 50~W. The CPU system has 64~GB of dual-channel DDR5-4800 memory with 76.8~GB/s peak bandwidth, while the Jetson has 64~GB of LPDDR5 memory with 204.8~GB/s peak bandwidth. \rev{Both platforms run the Bonsai and BitCPM models with dense attention and an FP16 KV cache; we additionally report the Jetson with an INT4 KV cache. In this comparison, the energy of \design includes HBM, whereas CPU energy is the RAPL package power and Jetson energy is measured on its module power rails.}
\item \textbf{ASIC and FPGA.} \design hardware is compared to the following ASIC and FPGA baselines: FIGLUT~\cite{park2025figlut}, TENET~\cite{huang2025tenet}, TeLLMe~\cite{qiao2025tellme}, MCBP~\cite{wang2025mcbp}, FACT~\cite{qin2023fact}, Energon~\cite{zhou2022energon}, and LUT-DLA~\cite{li2025lutdla}. \rev{\key{For FIGLUT, TeLLMe, TENET, and MCBP, we reproduce each linear PE in RTL using our 28-nm flow and scale it to its published width (e.g., 16 TENET cores of $32{\times}64$).} For MCBP, we map ternary weights onto its binary patterns with sequential positive/negative passes, which is more area- and energy-efficient than a direct ternary mapping. Designs without sparse attention use our dense attention datapath. TENET's sliding window (up to 1.5K tokens) and MCBP's BGPP apply within their published ranges; attention is dense beyond them. All share HBM2 and run at 1~GHz, except TENET at 500~MHz. \key{Energy includes on-chip and HBM dynamic energy; Table~\ref{tab:sota_area_efficiency} lists on-chip power only.}}
\end{itemize}

\noindent\textbf{Hardware Simulation.}
We implement \design with the configuration in Table~\ref{tab:hardware_config} using SystemVerilog and synthesize the design with Synopsys Design Compiler using a commercial 28-nm library. The area and energy of on-chip SRAM are estimated by CACTI~\cite{balasubramonian2017cacti}. \rev{A trace-driven cycle-level simulator, calibrated against per-stage RTL cycle counts, evaluates end-to-end performance. Its input traces record the pages selected by Top-$k$ and the tokens kept by Top-$p$ at each decode step of model inference. The simulator converts them into HBM2 reads, page-granular for metadata and K and token-granular for V, and charges them at a sustained HBM2 bandwidth that we validate by replaying the same address traces in Ramulator2~\cite{luo2024ramulator2}. HBM2 access energy is 4~pJ/bit~\cite{oconnor2017finegrained}.}
%% 具体数字可以不说，就说明 Ramulator2, trace 怎么来的

\begin{figure}[t]
    \centering
    \includegraphics[width=\linewidth]{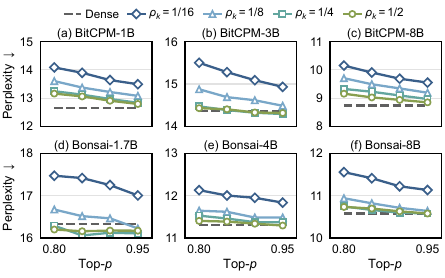}
    \caption{Perplexity under different Top-$k$ and Top-$p$ settings.}
    \label{fig:kp_grid}
\end{figure}

\begin{table}[t]
\centering
\caption{Quality comparison with dense attention.}
\label{tab:model_quality_longbench}
\scriptsize
\setlength{\tabcolsep}{2pt}
\resizebox{\linewidth}{!}{%
\begin{tabular}{llrrrrrrrrr}
\toprule
Model & Config & GovR & QMS & MNw & GSM8K & MBPP & IFE & CoQA & GMean & $|\Delta|$ \\
\midrule
BitCPM-1B & Dense & 25.9 & 21.2 & 25.3 & 50.8 & 42.4 & 64.1 & 54.1 & 37.4 & -- \\
 & Ours & 21.3 & 20.8 & 21.7 & 50.2 & 42.0 & 65.1 & 54.6 & 35.6 & 1.88 \\
\midrule
BitCPM-3B & Dense & 25.9 & 22.4 & 25.7 & 67.9 & 15.6 & 68.4 & 34.4 & 32.4 & -- \\
 & Ours & 23.8 & 22.3 & 23.9 & 67.6 & 16.4 & 68.4 & 35.7 & 32.0 & 0.36 \\
\midrule
BitCPM-8B & Dense & 27.1 & 23.6 & 25.6 & 62.5 & 56.2 & 69.1 & 73.1 & 43.4 & -- \\
 & Ours & 22.8 & 22.4 & 21.9 & 68.8 & 55.4 & 68.4 & 73.0 & 41.5 & 1.89 \\
\midrule
Bonsai-1.7B & Dense & 31.4 & 21.4 & 24.7 & 40.6 & 26.2 & 70.6 & 35.0 & 33.1 & -- \\
 & Ours & 28.3 & 20.5 & 19.7 & 39.7 & 26.8 & 70.8 & 34.6 & 31.4 & 1.75 \\
\midrule
Bonsai-4B & Dense & 32.0 & 20.1 & 23.3 & 79.8 & 47.2 & 74.5 & 44.8 & 40.8 & -- \\
 & Ours & 30.0 & 24.1 & 23.9 & 80.4 & 47.2 & 72.6 & 44.2 & 41.5 & 0.66 \\
\midrule
Bonsai-8B & Dense & 33.2 & 20.5 & 25.7 & 72.3 & 43.6 & 80.8 & 52.7 & 42.1 & -- \\
 & Ours & 27.9 & 22.0 & 19.2 & 72.9 & 43.4 & 81.0 & 53.2 & 39.9 & 2.21 \\
\bottomrule
\end{tabular}}
\par\vspace{3pt}
\begin{minipage}{\linewidth}
\footnotesize
GMean is the geometric mean of seven task scores; $|\Delta|$ is its absolute change. GSM8K uses strict EM, MBPP pass@1, IFE prompt-level strict accuracy, and CoQA F1. Short contexts disable Top-$k$.
\end{minipage}
\end{table}

\subsection{Algorithm Evaluation} \label{sec:eval_algorithm}

We evaluate the adaptive sparse attention algorithm from three perspectives: parameter sensitivity, quality across models and tasks, and reductions in computation and KV traffic.

\begin{figure}[t]
    \centering
    \includegraphics[width=\linewidth]{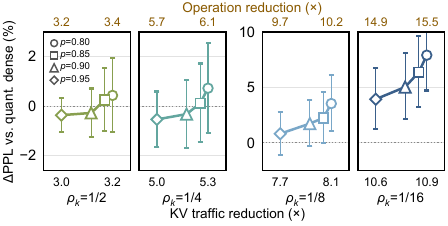}
    \caption{Quality--efficiency tradeoff of sparse attention for BitCPM-3B on WikiText-2 at 8K context length.}
    \label{fig:hyperparam_tradeoff}
\end{figure}

\begin{figure}[t]
    \centering
    \includegraphics[width=\linewidth]{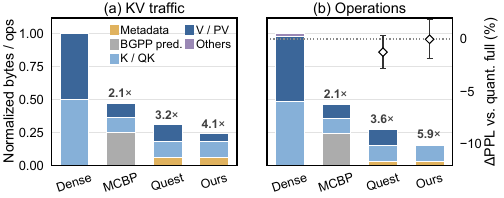}
    \caption{Reductions in KV traffic and operations under different attention schemes for BitCPM-3B at an 8K context length.}
    \label{fig:compute_traffic}
\end{figure}

\noindent\textbf{Sparse Attention Parameters.}
\figref{fig:kp_grid} sweeps the Top-$k$ page-retention ratio $\rho_k$ from $1/16$ to $1/2$ and the Top-$p$ threshold $p$ from $0.80$ to $0.95$. Across all six models, $\rho_k$ is the primary quality knob: reducing the page budget generally increases perplexity, while increasing $p$ tends to recover quality within each budget. At $\rho_k=1/4$ and $p=0.90$, four models remain within 1.3\% of their dense baselines, and the worst degradation among the remaining models is 4.0\%. We therefore adopt $\rho_k=1/4$ and $p=0.90$ as the default parameters.

\noindent\textbf{Quality across Models and Tasks.}
Table~\ref{tab:model_quality_longbench} \rev{shows the quality of the selected operating point across model scales and task types. Bonsai-4B slightly improves over its dense baseline, whereas the loss is larger on long-document summarization such as GovReport and MultiNews: one page set must serve all heads of a KV group, so pages needed by only a few heads may fall outside the budget, and Top-$p$ cannot recover pages excluded by Top-$k$. \key{Retrieval-style tasks are similarly sensitive, which is a limitation of the fixed page budget.}} Short contexts contain too few pages for page-level Top-$k$ to provide a useful tradeoff: pruning half of the pages causes a large quality loss, while retaining more pages leaves little KV traffic to save. We therefore disable Top-$k$ for short-context tasks but retain Top-$p$ where quality permits. 

\noindent\textbf{Computation and Data Traffic Reduction.}
\figref{fig:hyperparam_tradeoff} shows that $\rho_k$ determines the quality--efficiency tradeoff, while $p$ provides finer adjustment within each candidate fraction. A smaller $\rho_k$ yields larger traffic and operation reductions but sharply increases perplexity, whereas $\rho_k\geq1/4$ remains within the dense confidence band across the tested Top-$p$ thresholds. At the default $\rho_k=1/4$ and $p=0.90$, \figref{fig:compute_traffic} shows that the two stages provide complementary savings: Top-$k$ reduces K traffic, while Top-$p$ further reduces V traffic and PV computation. They together reduce KV traffic by $4.1\times$ and operations by $5.9\times$. Quest achieves only $3.2\times$ traffic reduction, and MCBP provides smaller savings after its predictor overhead is included~\cite{wang2025mcbp}. The Top-$p$ engine accounts for only 4.2\% of the remaining operations. \rev{Twilight~\cite{lin2026twilight} needs an extra low-precision K pass, which reads $2.1\times$ more KV bytes per head than \design at 8K ($1.02\times$ with an INT4 auxiliary cache).}

\subsection{Hardware Performance} \label{sec:eval_hardware}

\begin{figure}[t]
    \centering
    \includegraphics[width=\linewidth]{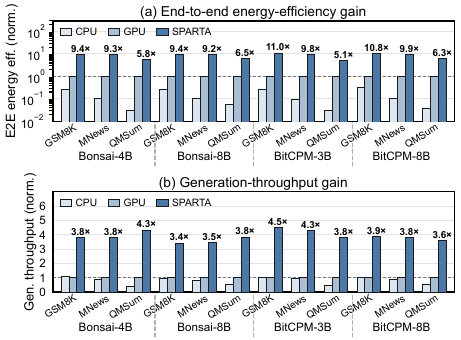}
    \caption{End-to-end energy efficiency and throughput.}
    \label{fig:comparison_cpu_gpu}
\end{figure}

\noindent\textbf{Comparison to CPU and GPU.}
\figref{fig:comparison_cpu_gpu} compares \design with the CPU and GPU baselines. \design consistently outperforms both commodity platforms across \rev{four} models. Compared with the GPU, \design improves geometric-mean generation throughput by \rev{$3.9\times$} and end-to-end energy efficiency by \rev{$8.3\times$}. The corresponding improvements over the CPU are \rev{$5.0\times$ and $81.7\times$}. \rev{With an INT4 KV cache, the Jetson becomes slower at long contexts because \texttt{llama.cpp} dequantizes the cache before attention, and the gains of \design rise to $4.8\times$ and $9.1\times$.} \rev{Over the CPU, the throughput gain of \design is largest on QMSum, the longest-context task, because more K/V accesses can be eliminated. Over the GPU, BitCPM gains least on QMSum because its two KV heads keep dense KV reads small.} \rev{Both baselines use dense attention, so these gains combine architectural and sparse-attention benefits.}
% , \design raises its throughput advantage over the GPU to $3.4\times$ and $3.7\times$, respectively; the corresponding gains over the CPU are $3.6\times$ and $7.4\times$. End-to-end energy efficiency remains $7.1$--$10.1\times$ higher than the GPU across all evaluated points because the dedicated datapath avoids the power overhead of a general-purpose platform.

\begin{figure}[t]
    \centering
    \includegraphics[width=\linewidth]{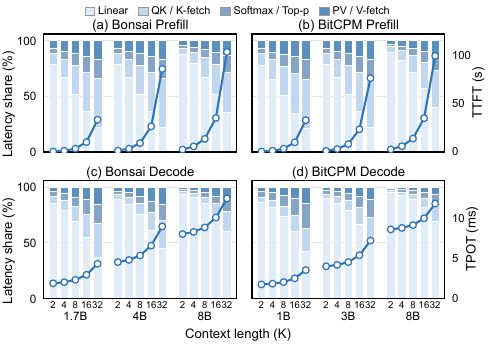}
    \caption{Latency breakdown by model and context length.}
    \label{fig:stage_breakdown}
\end{figure}

\noindent\textbf{Performance Breakdown.}
\figref{fig:stage_breakdown} shows \design's latency breakdown across linear layers and the three attention stages for different context lengths and models.

\noindent\emph{Prefill.} The latency bottleneck shifts from linear layers to attention as the context grows. Projection and FFN overhead grows linearly with prompt length, whereas dense attention grows quadratically. Consequently, the average linear-layer share falls from about 82\% at 2K to 28\% at 32K. At the same context length, larger models retain a higher linear-layer share because projection and FFN work grows with model width.

\begin{figure}[t]
    \centering
    \includegraphics[width=\linewidth]{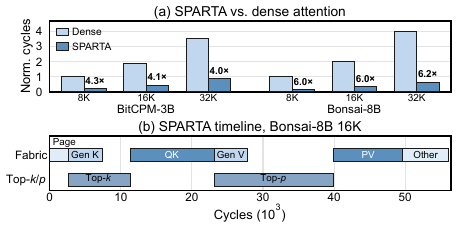}
    \caption{\rev{Decode attention speedup and stage overlap.}}
    \label{fig:stage_timeline}
\end{figure}

\noindent\emph{Decode.} Decode similarly shifts from linear-layer computation toward attention as the context grows. The linear work per generated token remains fixed for a given model, while attention must process an increasing number of cached tokens, reaching 31--63\% of TPOT at 32K. \rev{\figref{fig:stage_timeline}(a) shows that two-stage sparse attention is 4.0--6.2$\times$ faster than dense attention (normalized to each model's dense attention at 8K). \figref{fig:stage_timeline}(b) draws the stage overlap of \figref{fig:arch_overview} to scale: \key{Top-$k$ selection fully covers K generation except for Bonsai at 8K, and Top-$p$ selection covers V generation in all cases.}}

\begin{figure}[t]
    \centering
    \includegraphics[width=\linewidth]{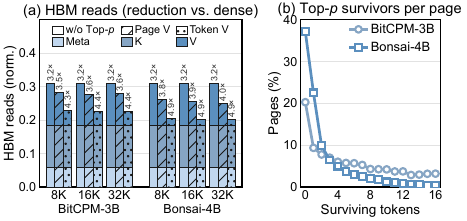}
    \caption{\rev{Decode HBM read traffic simulated with Ramulator2.}}
    \label{fig:hbm_granularity}
\end{figure}

\begin{figure}[!t]
    \centering
    \includegraphics[width=0.80\linewidth]{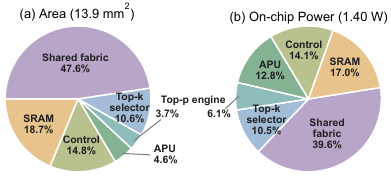}
    \caption{Area and average on-chip power breakdown.}
    \label{fig:area_power_breakdown}
\end{figure}

%% 32-byte 是怎么来的
\noindent\rev{\textbf{HBM Transfer Granularity.} \figref{fig:hbm_granularity} models decode KV reads in Ramulator2 as 32-byte HBM2 transactions (a 64-bit pseudo channel with burst length 4): metadata and K are fetched per selected page, while V is gathered per surviving token (Token V) as two transactions per 64-byte vector. \key{The transferred bytes are therefore only 1.4--1.7\% above the analytical bytes of the selected metadata and tokens.} As shown in \figref{fig:hbm_granularity}(a), \key{Top-$k$ reduces reads by $3.2\times$ at all lengths, and Top-$p$ raises the reduction to 4.3--4.4$\times$ for BitCPM-3B and $4.9\times$ for Bonsai-4B.} Only 20--27\% and 37--47\% of the selected pages are fully pruned (\figref{fig:hbm_granularity}(b)), so fetching whole V pages (Page V) would retain only 3.5--3.6$\times$ and 3.8--4.0$\times$.}

\begin{figure}[t]
    \centering
    \includegraphics[width=\linewidth]{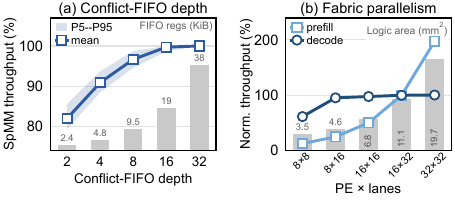}
    \caption{Fabric parameter sensitivity.}
    \label{fig:fabric_dse}
\end{figure}

\noindent\textbf{Area and Power.}
\figref{fig:area_power_breakdown} reports a total area of 13.9~mm$^2$ and average on-chip power of 1.40~W. The shared fabric accounts for 47.6\% of the area and 39.6\% of the power because it is replicated across eight compute slices and executes all linear and attention operations. In comparison, the Top-$k$ and Top-$p$ engines together occupy 14.3\% of the area and 16.6\% of the power, keeping the hardware cost of two-stage selection low.

\subsection{Design Space Exploration}
\label{sec:eval_design_params}

\begin{figure}[t]
    \centering
    \includegraphics[width=\linewidth]{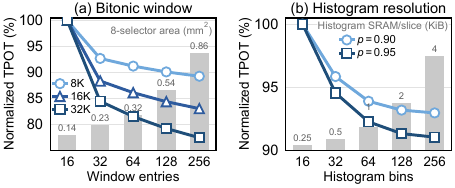}
    \caption{Selection-engine parameter sensitivity.}
    \label{fig:selection_dse}
\end{figure}

We sweep the conflict-FIFO depth, fabric parallelism, Top-$k$ window, and Top-$p$ histogram resolution.

\noindent\textbf{Conflict-FIFO Depth.}
A shallow FIFO frequently stalls SpMM when multiple nonzero weights target the same bank, whereas a deeper FIFO increases register cost. \figref{fig:fabric_dse}(a) shows that throughput rises quickly and then saturates. A depth of 8 reaches 96.7\% of the depth-32 throughput with only one quarter of the entries, so we set the conflict-FIFO depth to 8.

\noindent\textbf{Fabric Parallelism.}
\figref{fig:fabric_dse}(b) reveals different scaling behavior across inference phases. Prefill continues to benefit from additional lanes, while bandwidth-bound decode saturates much earlier. We select 16 PEs with 32 lanes: reducing the lanes to 16 halves prefill throughput, whereas increasing to 32 PEs adds 78\% logic area without reducing decode latency.

\noindent\textbf{Bitonic Top-$k$ Window.}
A larger sorting window reduces the number of merge iterations but rapidly increases comparator area. Relative to the 16-entry baseline in \figref{fig:selection_dse}(a), a 64-entry window reduces TPOT by 8.8\%, 13.9\%, and 18.5\% at sequence lengths of 8K, 16K, and 32K, respectively. Increasing the window to 128 entries provides only another 1.1--2.3 percentage points of reduction, while the selector area increases from 0.32 to 0.54~mm$^2$. We therefore use 64 entries.

\noindent\textbf{Histogram Resolution.}
More histogram bins improve the Top-$p$ threshold estimate but increase both threshold-search latency and SRAM cost. Relative to the 16-bin baseline in \figref{fig:selection_dse}(b), 64 bins reduce TPOT by 6.1\% at $p=0.90$ and 7.7\% at $p=0.95$. Increasing to 128 bins provides less than one additional percentage point of reduction while doubling the histogram SRAM from 1 to 2~KiB per compute slice. The 64-bin result is also statistically indistinguishable from exact Top-$p$ in the quality sweep, so we use 64 bins. \rev{\key{Replacing the histogram with Twilight's binary-search threshold on the same fabric slows attention by 2.0--3.0$\times$ and TPOT by 1.1--1.8$\times$.}}

\subsection{Ablation Studies} \label{sec:eval_ablation}

\begin{table}[t]
\centering
\caption{Sparse-attention quality ablation.}
\label{tab:algorithm_ablation}
\scriptsize
\setlength{\tabcolsep}{2.7pt}
\begin{tabular}{lllccccc}
  \toprule
  \textbf{Model} & \textbf{Data} & \textbf{Len} & \textbf{\boldmath $\rho_p$} & \textbf{Full} & \textbf{\boldmath $+$Top-$k$} & \textbf{\boldmath $+$Exact Top-$p$} & \textbf{\boldmath $+$Hist. Top-$p$} \\
  \midrule
  \multirow{4}{*}{BitCPM-3B}
    & \multirow{2}{*}{WT2}  & 8K   & 6.0\% & 14.37 & 14.19 & 14.34 & 14.36 \\
    &                       & 32K  & 5.0\% & 16.52 & 16.43 & 16.35 & 16.39 \\
    \cmidrule(l){2-8}
    & \multirow{2}{*}{PG19} & 8K   & 7.3\% & 16.93 & 17.49 & 17.64 & 17.54 \\
    &                       & 32K  & 6.5\% & 16.81 & 16.46 & 16.26 & 16.30 \\
  \midrule
  \multirow{4}{*}{Bonsai-4B}
    & \multirow{2}{*}{WT2}  & 8K   & 4.3\% & 11.31 & 11.36 & 11.42 & 11.46 \\
    &                       & 32K  & 2.9\% & 10.56 & 10.48 & 10.58 & 10.56 \\
    \cmidrule(l){2-8}
    & \multirow{2}{*}{PG19} & 8K   & 4.7\% & 16.62 & 16.73 & 16.96 & 16.88 \\
    &                       & 32K  & 3.3\% & 21.29 & 20.85 & 20.93 & 20.85 \\
  \midrule
  \multirow{4}{*}{BitCPM-8B}
    & \multirow{2}{*}{WT2}  & 8K   & 9.7\% & 8.74 & 8.82 & 9.03 & 9.05 \\
    &                       & 32K  & 8.2\% & 9.40 & 9.56 & 9.79 & 9.77 \\
    \cmidrule(l){2-8}
    & \multirow{2}{*}{PG19} & 8K   & 10.9\% & 11.75 & 11.86 & 12.08 & 12.08 \\
    &                       & 32K  & 10.1\% & 11.35 & 11.53 & 11.54 & 11.51 \\
  \bottomrule
\end{tabular}
\par\vspace{3pt}
\begin{minipage}{\linewidth}
\footnotesize
$\rho_p$ is the fraction of Top-$k$ candidates retained by Top-$p$.
\end{minipage}
\end{table}

\noindent\textbf{Algorithmic Ablation.}
Starting from full attention, Table~\ref{tab:algorithm_ablation} cumulatively enables page-level Top-$k$, exact Top-$p$, and the 64-bin histogram Top-$p$. Sparse selection preserves perplexity across models, datasets, and context lengths: the final configuration differs from full attention by at most 0.61 PPL. The histogram engine also closely matches exact Top-$p$, changing perplexity by at most 0.10. The retained candidate fraction $\rho_p$ is only 2.9--10.9\%, and it decreases from 8K to 32K. The histogram approximation therefore adds negligible quality loss while retaining fewer tokens at longer contexts.

\begin{figure}[!t]
    \centering
    \includegraphics[width=\linewidth]{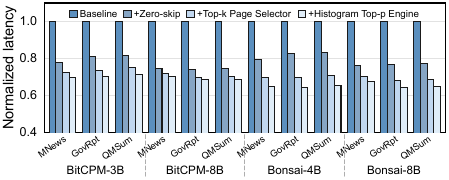}
    \caption{Cumulative end-to-end latency ablation.}
    \label{fig:hardware_ablation}
\end{figure}

\noindent\textbf{Hardware Ablation.}
We evaluate every configuration on the same prompts with a fixed output length of 512 tokens and report normalized end-to-end latency. \figref{fig:hardware_ablation} shows that each mechanism consistently reduces latency across all model--dataset pairs. Zero-weight skipping provides the largest initial gain by eliminating ineffective ternary computation. The Top-$k$ page selector then reduces K-cache accesses, and the histogram Top-$p$ engine further reduces V traffic and PV computation. At the geometric mean, these three steps reduce latency by 22\%, 9\%, and 5\% relative to the preceding configuration, yielding a cumulative $1.48\times$ speedup. Zero-weight skipping mainly shortens the compute-bound prefill (TTFT $-29\%$), whereas Top-$k$ and Top-$p$ act only in decode, where they reduce TPOT by 26\% and 16\%.

\subsection{Comparison with SOTA LLM Accelerators} \label{sec:eval_sota}

\begin{figure}[t]
    \centering
    \includegraphics[width=\linewidth]{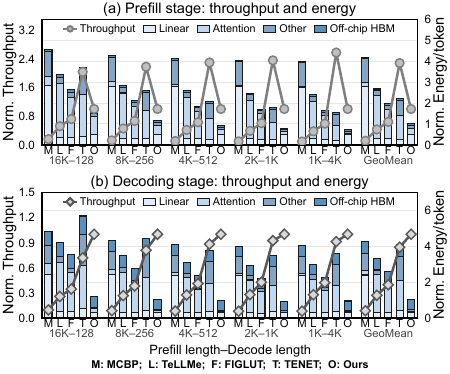}
    \caption{Throughput and energy efficiency comparison.}
    \label{fig:sota_iso_resource}
\end{figure}
%% off-chip dynamic -> DRAM/HBM

\noindent\textbf{Throughput and Energy.}
\figref{fig:sota_iso_resource} compares \design with FIGLUT, TeLLMe, TENET, and MCBP~\cite{park2025figlut,qiao2025tellme,huang2025tenet,wang2025mcbp} under different prefill--decode lengths. Throughput and energy per token are both normalized to the geometric mean of \design within each phase.

\noindent\emph{Prefill.} Prefill remains compute-bound across the evaluated prompt lengths, so the relative throughput ordering is stable. At the geometric mean, \design achieves $1.6\times$, $2.3\times$, and $8.8\times$ higher throughput than FIGLUT, TeLLMe, and MCBP, and 2.0--4.2$\times$ higher energy efficiency than all four designs. \rev{Bitmap-compressed zero-weight skipping explains the gain over FIGLUT, while the separate dense-attention engines of FIGLUT and TeLLMe consume 73--76\% of their prefill energy. TENET occupies about $7\times$ the chip area of \design; at equal area, \design is $3.1\times$ faster.}

\noindent\emph{Decode.} The gains over the dense-attention baselines grow as the KV cache expands because two-stage selection reduces both K and V accesses. At the geometric mean, \design achieves $2.5\times$, $3.7\times$, $1.2\times$, and $12.0\times$ higher throughput than FIGLUT, TeLLMe, TENET, and MCBP, and $2.6\times$, $3.3\times$, $4.1\times$, and $4.2\times$ higher energy efficiency. The gain over TENET rises from $1.1\times$ at short prompts to $1.4\times$ at 16K, where most of the context lies beyond its 1.5K-token window. \rev{MCBP prunes attention with its BGPP predictor, so its gap stays nearly constant (10.6--12.6$\times$); 56\% of its decode energy goes to linear layers and only 6\% to attention.}

\begin{table}[t]
    \centering
    \caption{Summary and comparison with SOTA works.}
    \label{tab:sota_area_efficiency}
    % Auto-generated by scripts/plot_sota_iso_resource.py.
\fontsize{7.5}{8.2}\selectfont
\setlength{\tabcolsep}{1.9pt}%
\renewcommand{\arraystretch}{1.08}
\begin{tabular}{@{}lrrrrrr@{}}
\toprule
\textbf{Design} & \textbf{Tech.} & \textbf{Freq.} & \textbf{Area} & \textbf{Thpt.} & \textbf{Energy eff.} & \textbf{Area eff.} \\
 & \textbf{[nm]} & \textbf{[GHz]} & \textbf{[mm$^2$]} & \textbf{[GOPS]} & \textbf{[GOPS/W]} & \textbf{[GOPS/mm$^2$]} \\
\midrule
\multicolumn{7}{l}{\emph{Sparse-attention accelerators}} \\
Energon~\cite{zhou2022energon}$^{a}$ & 45 & 1.0 & 1.63 & 1,153 & 7,206 & 709 \\
FACT~\cite{qin2023fact} & 28 & 0.5 & 6.03 & 928 & 2,754 & 154 \\
\addlinespace[1pt]
\multicolumn{7}{l}{\emph{Linear-layer accelerators}} \\
FIGLUT~\cite{park2025figlut}$^{b}$ & 28 & 0.1 & 7.34 & 140 & 470 & 19.1 \\
LUT-DLA~\cite{li2025lutdla}$^{c}$ & 28 & 0.3 & 3.64 & 2,765 & 5,600 & 759 \\
\addlinespace[1pt]
\multicolumn{7}{l}{\emph{End-to-end LLM accelerators}} \\
TENET~\cite{huang2025tenet}$^{d}$ & 28 & 0.5 & 91.0 & 40,960 & 7,186 & 450 \\
\rowcolor{Purple!10}\textbf{Ours}$^{e}$ & \textbf{28} & \textbf{1.0} & \textbf{13.9} & \textbf{13,653} & \textbf{9,752} & \textbf{982} \\
\bottomrule
\end{tabular}
\par\vspace{3pt}
\parbox{\linewidth}{\footnotesize $^{a}$Scaled to 28~nm. $^{b}$Our reproduction. $^{c}$Design~3. $^{d}$16 STL + 4 HP cores of $32{\times}64$. $^{e}$40\% weight sparsity.}

\end{table}

\noindent\textbf{Hardware Efficiency.}
Table~\ref{tab:sota_area_efficiency} compares the reported metrics by acceleration scope. The sparse-attention and linear-layer accelerators are optimized for partial inference pipelines, whereas \design leverages the sparsity in both linear and attention layers across prefill and decode phases. \design achieves $1.4$--$20.7\times$ higher energy efficiency and $1.3$--$51.4\times$ higher area efficiency than the compared designs. \rev{\key{TENET pairs dedicated ternary linear (STL) cores with separate attention (HP) cores and thus represents heterogeneous designs; \design achieves $1.4\times$ its energy efficiency and $2.2\times$ its area efficiency.}}

%%%%%%%%%%%%%%%%%%%%%%%%%%%%%%%%%%%%%%%%%%%%%%%%%%%%%%%%%%%%%%%%%%%%%%%%%%%%%%%%%%%%%%%%%%
%%%%%%%%%%%%%%%%%%%%%%%%%%%%%%%%%%%%%%%%%%%%%%%%%%%%%%%%%%%%%%%%%%%%%%%%%%%%%%%%%%%%%%%%%%
\section{Related Work}
\label{sec:related_work}

\noindent\textbf{Ternary and Low-Bit Acceleration.}
\rev{Bit-Pragmatic exploits activation bit sparsity, Bit-Tactical schedules nonzero weights with activation matching, and Laconic pairs effectual bits from both operands~\cite{albericio2017bit,lascorz2019bittactical,sharify2019laconic}. \key{\design adopts their zero-bit skipping for the Q/P operand, but each set bit drives a shifted select-add update of a whole K/V vector on the same datapath that runs ternary weights.}} FIGLUT/LUT Tensor Core use lookup tables; MCBP reuses bit-slice patterns~\cite{park2025figlut,mo2025lut,wang2025mcbp}. TENET, TeLLMe, and TOM specialize ternary linear execution~\cite{huang2025tenet,qiao2025tellme,guan2026tom}. \rev{BitMoD supports mixed datatypes with bit-serial exponent alignment, normalization, and dequantization~\cite{chen2025bitmod}; \design keeps integer operands, so its update needs only shifts and additions. VitaLLM partially reuses BoothFlex, but TINT cannot execute multi-bit attention, limiting resource reallocation as workloads change~\cite{lin2026vitallm}. \design shares integer select-add updates across ternary linear and Q/P-expanded attention.}

\noindent\textbf{KV-Cache Compression and Sparse Attention.}
Existing schemes reduce KV precision~\cite{lin2024qserve,liu2024kivi,hooper2024kvquant}; eviction methods retain compact subsets~\cite{xiao2024efficient,zhang2023h2o,li2024snapkv,adnan2024keyformer,cai2024pyramidkv}; query-time methods predict relevant entries~\cite{ribar2023sparq,singhania2024loki,liu2026retrievalattention,gao2024seerattention}. Accelerators approximate or predict sparse QK work~\cite{ham2020A3,ham2021elsa,wang2021spatten,lu2021sanger,qu2022dota,zhou2022energon,qin2023fact,wang2024sofa}; PADE fuses sparsity discovery with execution~\cite{wang2026pade}. MInference/LServe sparsify prefill or both phases; Quest, Double Sparsity, and Twilight provide page Top-$k$, channel/token sparsity, and hierarchical Top-$p$~\cite{jiang2024minference,yang2025lserve,tang2024quest,yang2024post,lin2026twilight}. \rev{VitaLLM selects token Top-$k$ from leading-one approximations~\cite{lin2026vitallm}; \design uses metadata-based page Top-$k$ followed by exact candidate QK and token Top-$p$.}

\noindent\textbf{Prefill--Decode and Attention Dataflows.}
Serving systems split phases or schedule chunked prefill~\cite{patel2024splitwise,zhong2024distserve,agrawal2024taming}. FlexGen/PagedAttention manage memory; FlashInfer/FlashAttention optimize attention kernels and I/O~\cite{sheng2023flexgen,kwon2023efficient,ye2025flashinfer,dao2022flashattention,dao2023flashattention,hong2311flashdecoding++}. FLAT/FuseMax optimize spatial QK--Softmax--PV~\cite{kao2023flat,nayak2024fusemax}. \design instead time-multiplexes prefill SpMM, decode SpMV, and sparse QK/PV on one fabric.

%%%%%%%%%%%%%%%%%%%%%%%%%%%%%%%%%%%%%%%%%%%%%%%%%%%%%%%%%%%%%%%%%%%%%%%%%%%%%%%%%%%%%%%%%%
%%%%%%%%%%%%%%%%%%%%%%%%%%%%%%%%%%%%%%%%%%%%%%%%%%%%%%%%%%%%%%%%%%%%%%%%%%%%%%%%%%%%%%%%%%
\section{Conclusion}
\label{sec:conclusion}
This paper presented \design, a sparse ternary LLM accelerator for long-context edge inference. \design unifies ternary linear computation and adaptive sparse attention on a shared compute fabric, jointly exploiting zero-weight, bit-level, and query-dependent KV sparsity across both prefill and decode. Its two-stage adaptive attention further reduces KV-cache traffic with a quality--traffic tradeoff that depends on the task. Experimental results demonstrate that \design significantly improves throughput and energy efficiency over CPUs, GPUs, and SOTA LLM accelerators, enabling efficient long-context inference for ternary LLMs on edge devices.

%%%%%%% -- PAPER CONTENT ENDS -- %%%%%%%%

%%%%%%%%% -- BIB STYLE AND FILE -- %%%%%%%%
\clearpage

\bibliographystyle{unsrtnat}
\small
\bibliography{refs}
%%%%%%%%%%%%%%%%%%%%%%%%%%%%%%%%%%%%

% Required appendix on AI use (after references, not counted in the page limit).
\normalsize
\section{AI Use}
The authors designed the architecture and experiments and are responsible for all content of this manuscript. Claude (Anthropic) and Codex (OpenAI) assisted in drafting and editing the text and in writing analysis and plotting scripts. The authors reviewed, revised, and verified all AI-assisted content before including it in the manuscript.

\end{document}